\documentclass[10pt,twocolumn]{article}
\usepackage[T1]{fontenc}
\usepackage{csquotes}
\usepackage{comment}
\usepackage{epigraph}

\usepackage{graphicx}
 
\usepackage{xcolor}
\usepackage{xspace}

\usepackage[compress]{natbib}
\usepackage[hidelinks,colorlinks=false]{hyperref}
\definecolor{oxblue}{rgb}{0,0.1294117647,0.27843137254} 
\definecolor{forestgreen}{HTML}{008000}
 \hypersetup{
 colorlinks=true,
 linkcolor=black,
 filecolor=black,
 citecolor=forestgreen, 
 urlcolor=black
 }

\usepackage{fullpage} 
\usepackage[switch]{lineno}
\usepackage{array}
\usepackage{hhline}
\usepackage{amssymb}
\usepackage{mathtools}
\usepackage{amsmath} 
\usepackage{authblk} 

\usepackage[mediumspace,mediumqspace,squaren]{SIunits}

\DeclareRobustCommand{\cor}[1]{#1}

\usepackage[capitalise,nameinlink,sort&compress]{cleveref}
\crefname{equation}{}{}
\creflabelformat{equation}{#2#1#3}
\crefmultiformat{equation}{#2#1#3}{, #2#1#3}{, #2#1#3}{, #2#1#3}
\renewcommand\eqref[1]{(\cref{#1})}

\newcommand{\matr}[1]{\mathbf{#1}}
\renewcommand{\vec}[1]{\mathbf{#1}}

\newcommand{\pp}[1]{\left({#1}\right)}

\newcommand{\eqdef}{\vcentcolon =}

\newcommand{\linepdiff}[2]{{\partial{#1}}/{\partial{#2}}}

\newcommand{\ramp}[1]{\pp{#1}_+} 

\begin{document}

\title{Towards an active matter theory of plant morphogenesis}

\author[1]{Hadrien Oliveri\thanks{\href{mailto:holiveri@mpipz.mpg.de}{holiveri@mpipz.mpg.de}}}
\author[2]{Christophe Godin\thanks{\href{mailto:christophe.godin@inria.fr}{christophe.godin@inria.fr}}}
\author[2,3]{Ibrahim Cheddadi\thanks{\href{mailto:ibrahim.cheddadi@univ-grenoble-alpes.fr}{ibrahim.cheddadi@univ-grenoble-alpes.fr}}}

\affil[1]{Max Planck Institute for Plant Breeding Research, 
Cologne 50829, Germany}
\affil[2]{Laboratoire de Reproduction et Développement des Plantes, Université de Lyon, ENS de Lyon, UCBL, INRAE, CNRS, Inria, Lyon 69364, France}
\affil[3]{Université Grenoble Alpes, CNRS, Grenoble INP, TIMC, Grenoble 38000, France}

\maketitle 

\begin{abstract}
Plant morphogenesis relies on dynamic growth deformations at the cell and tissue scales driven by osmotic fluxes. A mechanistic understanding of this phenomenon demands a physical framework that integrates cell imbibition, tissue mechanics, and water fluxes, as well as their biophysical and molecular regulations, within a theory of plant active matter capturing the open-system and out-of-equilibrium properties of tissues. Building on historical insights into growth geometry, physics, and mechanics, combined with recent experimental results, we outline the key challenges in modelling plant growth and propose steps towards a unified physical theory of plant morphogenesis, in which biological regulation,  mechanical forces, and water fluxes interact to shape biological form through the fundamental principles of living matter.
\end{abstract}


\epigraph{La vie d'une plante se confond avec sa croissance.}{Francis Hall\'e, \textit{\'Eloge de la plante}, 1999}
 
\section{Introduction} 

\textit{Morphogenesis} is the biological process through which the form of a
cell, a tissue, or an organism is established.
In general, shape change occurs through a 
set of stereotypical, fundamental deformations driven by cell and organismal mechanics, e.g., in animal tissues, bending, tissue flow, or growth, which are organized in space and time. Tremendous progress in genetics and cellular biochemistry initially led to the doctrine that the morphogenetic information underlying this organization is deterministically
encoded by genes, which encapsulate the developmental program. This view was reinforced by the discovery of
so-called \textit{master genes}~\citep{halder1995}, which mediate organ formation.
However, the all-in-genes perspective neglects the emergent nature of morphogenetic processes and the multiphysical, multiscale, and nonlinear feedbacks that control the emergence of form, including the expression of genes themselves. Accordingly, the idea of genes forming the ``blueprint of life'' has fallen out of favour~\citep{noble2024s} and, instead, a large body of work has examined morphogenesis as a self-organized physical phenomenon, e.g., through Turing-type chemical instabilities in morphogen fields~\citep{murray2mathbiol};
 elastic instabilities in growing soft matter~\citep{Goriely2017,ben2025wrinkles}; or chemomechanical couplings in 
 the cytoplasm leading to spontaneous phase separation, e.g., in cell division~\citep{PhysRevLett.123.188101}.

\textit{Active matter} refers generically to any matter that takes energy
(e.g. chemical energy, heat)
from its environment to perform work (e.g. to move or deform). Its study has traditionally
focused on contractile and fluid-like, motile systems,
such as flocks~\citep{Toner_2024} 
or active
gels, such as actomyosin networks, with a wealth of applications to animals or bacteria
\citep{marchetti2013hydrodynamics,julicher2018hydrodynamic,saw2018biological}.
This description 
has permitted considerable advances in the study of animal morphogenesis, putting forth physically grounded theories of living matter. 
This approach has helped show that not only patterns of gene expression,
but also mechanics, physics, and geometry itself may convey morphogenetic information and
define the characteristic time and length scales controlling
morphogenesis~\citep{collinet2021}.

Plants differ fundamentally from animals in that their cells are mostly non-contractile and embedded in a rigid matrix of cellulose---the cell wall---which prevents their
migration. Hence, plant morphogenesis relies essentially on the addition of material through growth and cell division. 
The kinematics of growth has been extensively measured,
at organ~\citep{SILK1979481} or cellular scale, e.g., in the \textit{shoot apical
meristem}~\citep{kwiatkowska2003growth}. To explain the origin of the observed growth kinematics,~\cite{coen_genetics_2004} posited a direct causal link between gene-expression patterns and specific descriptors of the expansion rate. Using this paradigm in simulations, they demonstrated how complex forms can emerge by prescribing the spatial and temporal distributions of these parameters, thereby highlighting the essential role of genetically regulated growth and patterning in plant morphogenesis. Indeed, molecular regulatory networks and gene expression patterns have been the chief candidates for the control of morphogenesis.

But besides chemical patterning, 
plant growth is an active, self-organized, physical, and mechanical process, subject to complex regulation pathways and
physical couplings. Cells grow by absorbing water from their
surroundings, a hydromechanical phenomenon powered by high osmolarity,
actively maintained within the cells. Water entry generates
hydrostatic (\textit{turgor}) pressure in the cell, balanced by
mechanical tension within the cell walls. Simultaneously, high tensions cause the wall
to yield and expand irreversibly~\citep{cosgrove2005growth,ali2014physical}.
From a thermodynamic standpoint, the maintenance of cell chemical energy enables continuous growth by powering osmotic fluxes. This energy is then dissipated
through mainly two processes: (i) wall yield, remodelling, and
synthesis mediated by cell wall tension, and (ii) water
transport towards growing regions through various pathways---typically via cell-to-cell membrane connections (symplastic pathway) or in spaces surrounding cells (apoplastic pathway).

Historically, the cell wall has attracted considerable interest in the plant biology community, and its structure and molecular mechanics are now relatively well characterized. The cell wall is composed of a complex composite material consisting of cellulose fibres---microfibrils and hemicellulose---embedded within a pectin matrix~\citep{cosgrove_wall_2001}. Directionally-biased microfibril alignment confers anisotropic mechanical properties to the cell wall, limiting growth in the direction of these fibres. In turn, cellulose microfibrils are deposited by cellulose synthases whose trajectories are guided by cortical microtubules at the inner face of the plasma membrane~\citep{paredez2006visualization}. Through this coupling, cells regulate growth by modulating the mechanical properties of their walls through microtubule alignment. A mechanism for this regulation was hypothesised by \cite{hamant_developmental_2008}, who proposed a feedback loop between mechanical stresses within the cell wall and the organization of cortical microtubules. Although the underlying molecular basis for this coupling remains unclear, the hypothesis successfully accounts for the coordinated microtubule arrangements observed in epidermal cells and for characteristic patterns of apical morphogenesis.

Although more marginally studied in the context of morphogenesis, water plays a central role in plant development, and water transport and plant-water relations have been an important field of research as well~\citep{kramer1995water}.
Yet, in most growth studies, water is considered to be a non-limiting source of mechanical
work available to deform the cell walls and trigger their expansion through pressure forces. 
Consequently, most mechanical models of plant growth have focused on the
cell wall mechanical properties~\citep[see, e.g.,][]{boudon_computational_2015},
and turgor pressure has been treated as
a constant parameter. However, the relation between turgor and growth
is not straightforward in general~\citep{ali2023revisiting}. 
For instance, recent experimental and theoretical works have shown that cell pressure can be spatially heterogeneous across the shoot apex~\citep{long2020cellular}, 
with either negative or positive correlation with cell growth
rate. This dependence could be described parsimoniously with a cellular
model coupling water fluxes, wall mechanics, and growth~\citep{cheddadi2019coupling}. More generally, the simple fact that water has to be transported to account for the gain in volume corresponding to
growth reveals that a growing region, in essence, acts as a water
sink~\citep{cheddadi2019coupling, oliveri2025field}. Such a sink could inhibit growth in its neighbourhood, providing then a lateral inhibition mechanism, akin to molecular inhibition fields identified in phyllotaxis~\citep{douady1992}.
Such hydraulic inhibition was recently confirmed with the
experimental observation of shrinking cells in the shoot apex at the
boundary of growing primordia~\citep{alonso2024water}, which was also shown to contribute to defining the cellular identity of the boundary domain. 

Physics-based modelling of plant living matter is essential to understanding how these couplings shape plants. As in animals, the time and length scales
of morphogenetic patterns are defined by physical interactions. 
Here, we develop the idea that a theoretical description of
both tissue mechanics and water transport is essential to achieving sound physics-based modelling of plant growth. We review the history and development of growth studies and their
applications to plant development. 
By re-examining key concepts such as turgor pressure through the lens
 of biophysical principles, we aim to offer perspectives on an
 active-matter view of plant tissue---as a complex system
 dynamically regulated by coupled hydraulic, mechanical, and chemical
 interactions.

\section{Morphometrics: measuring form\label{morphology}}

\textit{What is growth?} In the introduction of his magnum opus \textit{On Growth and Form}, D'Arcy Thompson already grappled with this question: 
\blockquote[\citep{thompson1917arcy}]{While growth is a somewhat vague word for a complex matter, which may depend on various things, from simple imbibition of water to the
complicated results of the chemistry of nutrition, it deserves to be studied in relation to form.}
Thus, while growth is the result of complex mechanisms, its substance as a physical phenomenon seems at first to elude a general definition, and early authors set forth the intuitive idea of growth as a \textit{change in form}, highlighting a description of form in terms of specific mathematical functions~\citep{ambrosi2011perspectives}. Thompson explores various mathematical concepts of \textit{morphology} (a term which he attributes to Goethe; p. 719). In Chap. 3, he discusses the fundamental notion of \textit{rate of growth}. In Chap. 11--13, he reflects on the occurrence of spiral and helicoidal geometries in animals, e.g. shells, and plants. In Chap. 17, he presents his \textit{theory of transformation} which maps the forms of related species onto one another through smooth deformations of a Cartesian grid. 

\textit{Morphometrics}, the measurement of living forms and their change, is an old problem in science---How quickly does a human grow from a baby to an adult? How tall and wide does a tree grow? What defines a normal body shape? How do the relative proportions of body parts change during development? 
Which part of a leaf or a hand grows the fastest? Or how can one define a \textit{rate of growth}? 

These questions led to the development of \textit{allometry}---\blockquote[\citep{gayon2000history}]{the changes in relative dimensions of parts of an organism that are correlated with changes in overall size}---a reflection having origins in the work of Galileo and the anatomists 
Cuvier and 
Dubois, and synthesized in the 1920--30s by 
Huxley and 
Teissier~\citep{huxley1924constant,huxley1993problems,huxley1936terminology}. In the context of morphogenesis, these ideas gave rise to \textit{ontogenetic} allometry, which follows from the understanding that complex form requires different body parts of an organism to grow at different rates---a property termed \textit{heterogonic growth} by \cite{pezard1918conditionnement}. The idea then is to characterize the scaling relationship between the size of a body ($x$), and that of a subregion ($y$), in the form of a power law $y\sim x^\beta$---an approach employed by \cite{avery1933structure} in a classic study on tobacco leaves. 

This type of early characterization of growth nevertheless presents a number of evident mathematical shortcomings that hinder its physical interpretation~\citep{needham1934chemical,kavanagh1942mathematical,Goriely2017}. Instead, authors have turned to more robust 
concepts of continuum kinematics, 
 tracing their origins to the development of modern elasticity theory in the 19\textsuperscript{th} century. 
 These concepts resurfaced in the plant biology community, e.g. as the \textit{relative elemental growth rate} in the context of roots~\citep{erickson1956elemental}, or the \textit{elemental growth-rate in volume per unit volume} to measure growth of leaves~\citep{richards1943analysis}. A generalised view was further developed by \cite{skalak1982analytical,hejnowicz1984growth} with the introduction of various \textit{growth tensors}~\citep[see also][]{silk1984quantitative}, synthesizing the earlier intuition of Thompson: 
\blockquote[{\citep[][Chap. 3]{thompson1917arcy}}]{
The form of an animal\footnote{Thompson atoned for this early lapse of plant blindness by adopting the more inclusive term ``organism'' in the 1942 second edition.} 
is determined by its specific rate of growth in various directions;
accordingly, the phenomenon of rate of growth deserves to be
studied as a necessary preliminary to the theoretical study of
form.}
 
\begin{figure}
 \centering
\includegraphics[width=\linewidth]{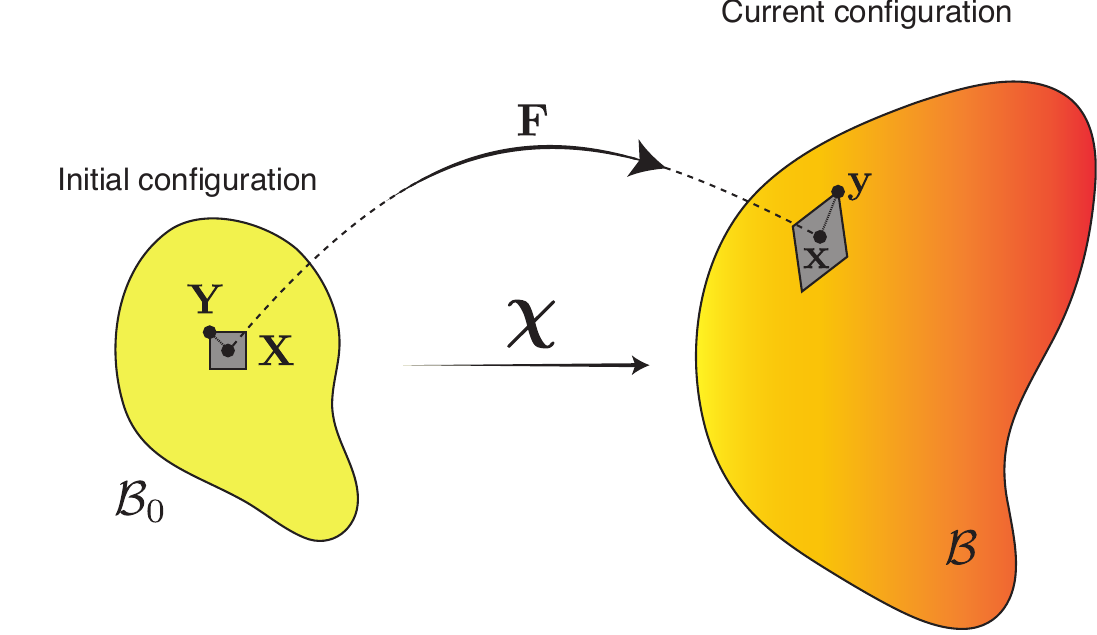}
 \caption{Deformation of the initial configuration $\mathcal B_0$ by the smooth map $\chi$ to the current configuration $\mathcal B$. The tensor $\matr F$ is the \textit{gradient of deformation tensor}.}
 \label{fig:geometry}
\end{figure}

The \textit{deformation gradient tensor} is a central concept in the mathematical study of body shape changes, characterizing the linear transformation undergone by \textit{infinitesimal} volumes within the solid. More precisely, let $\vec X$ and $\vec Y = \vec X+\Delta \vec X $ be a pair of material points located in the growing body at an initial time $t=0$. After growth, at a given time $t\geq 0$, these two material points have moved to two new positions in space, 
$\vec x$ and $\vec y = \vec x + \Delta \vec x$. 
The tracking of material points in time is represented by a mapping $\chi$, such that $\vec x = \chi \pp{\vec X, t}$ and $\vec y = \chi \pp{\vec Y, t}$. If $\vec X$ and $\vec Y$ are close to each other (in the sense that $\Delta \vec X \rightarrow \vec 0$), 
the two vectors $\Delta \vec X$ and $\Delta \vec x$ are linked through the deformation gradient $\matr F (\vec X,t)$ at $\vec X$ via the relation
\begin{equation}
\Delta \vec x \approx \matr F(\vec X, t) \Delta \vec X 
\end{equation}
(\cref{fig:geometry}).
Here $\matr F (\vec X,t) = \linepdiff{\chi}{\vec X}$ is the differential of $\chi$ w.r.t. the material position $\vec X$, which describes the integrated local expansion and rotation of material lines around the point $\vec X$ of interest, between times zero and $t$~\citep[cf.][for details]{holzapfel2000nonlinear}. In three-dimensional space, $\matr F$ is a $3\times3$ tensor. In one dimension, it reduces to the derivative $ F(X,t) = \linepdiff{\chi}{X} \approx \Delta x / \Delta X$. 

By differentiating $\matr F$ w.r.t. time, one obtains a \textit{rate of deformation tensor}, defined as $\matr L \eqdef \matr {\dot F}\matr F^{-1}$~\citep{holzapfel2000nonlinear}, with the overdot denoting differentiation w.r.t. time $t$. Indeed, this definition can be identified with the gradient of the spatial velocity field $\vec v (\vec x,t)$ (i.e., the instantaneous velocity of the material at a location $\vec x$) as $\matr L = \linepdiff{\vec v}{\vec x}$, 
providing a measure of how nearby points move relative to one another in physical space. 

The widespread adoption and success of continuum kinematics concepts in the context of biological growth are not entirely surprising:
\blockquote[{\citep{erickson1980}}]{
Like a flame or the wake of a boat, the form of a plant changes slowly, but the components are in continual flux. The motions of the components can therefore be analysed in terms of fluid flow.}
Following this analogy, empirical maps of strain rate and vorticity (rate of local rotation) have been constructed, e.g. in leaves~\citep{richards1943analysis,SILK1979481,wolf1986quantitative,Rolland-Lagan.2005,alim2016leaf,derr2018fluttering}. 
Progress in microscopy and image analysis further enabled quantitative kinematic analysis at the cell scale, now a primary means of investigation in modern plant developmental biology~\citep{dumais2002,kwiatkowska2003growth,de2015morphographx}.
These various works have established the importance of quantifying form to understand its origin. 
However, the sole description of geometry and kinematics lacks the explanatory power of \textit{scientific models}, which seek to capture phenomena in light of their causes, seeking not only to describe observed changes in form but also to account for their origin.

The origin of form is a longstanding problem in biology. 
Against the current of the Darwinian revolution, \textit{On Growth and Form} revisits the old concept of form through the lens of material sciences, interpreting growing bodies as `diagrams of forces'. 
\cite{richards1943analysis} later articulated this view, noting that:\blockquote[]{Geometrical change alone [...] may not give a completely satisfactory picture of the underlying growth activity. The change in size at a given point is due to both the functional activity of the cells located there and to the forces of stretch or compression exerted by the adjacent material.}
This perspective has motivated the study of morphogenesis through physical and mechanical principles. In parallel, another tradition in developmental biology has approached morphogenesis from the standpoint of genes, their evolution, and their expression. In the next few sections, we outline how these two views combine into a theory of form.

\section{Morphogenetics: the chemical basis of morphogenesis\label{morphogenetics}}

A chief approach to elucidating the mechanisms of morphogenesis in plants has centred on the study of chemical fields, e.g. genes and hormones, viewed as determinants of growth. Prior to modern genetics and cell biochemistry, such a perspective appeared in botany in classical studies of tropism.
In the 1870s, Charles Darwin and his son Francis explored phototropism, the bending of shoots towards light~\citep{darwin1880power}. They proposed that a `substance' propagated from the photosensitive apex of the shoot down to the growing regions, where bending occurs. This research culminated with the hormonal theory of tropism established based on work by Boysen Jensen, Cholodny, Thimann and Went at the beginning of the 20\textsuperscript{th} \cor{century}, and the isolation of \textit{auxin} by K\"ogl and Haagen-Smit~\citep{arteca1996plant,whippo2006phototropism}. Auxin is now associated with virtually all morphogenetic processes. In tropism, shoot curvature arises from differential growth of the tissue stimulated by heterogeneous auxin distribution~\citep{muday2001auxins,moulton2020multiscale}. 
The general mechanisms of spatial organization of auxin have become a crucial area of study in plant development, e.g. in leaves~\citep{scarpella2010control,bilsborough2011model} and roots~\citep{grieneisen2007auxin,Band.2014}, or in phyllotaxis~\citep{vernoux_auxin_2010,traas_phyllotaxis_2013,Vernoux.2021}.

Generally, a defining property of life lies in the ability of chemical determinants of development to distribute heterogeneously in space to generate complex forms. How chemical patterning emerges is a key problem in development: How do cheetahs get their dots? How does the Drosophila embryo segment itself? How do Fibonacci spirals form at the surface of a fir cone? How do we get five fingers on each hand and thirty-two teeth in our mouths?

 In his celebrated paper \textit{The chemical basis of morphogenesis}, Alan Turing proposed a theoretical framework for studying such questions mathematically~\citep{turing1952chemical}. Turing posited a generic system of partial differential equations describing the concentrations of $n\geq 2$ chemicals interacting locally, and diffusing in the domain of the tissue. Remarkably, under certain theoretical conditions, these species may behave in such a way that the uniform state (which in the absence of diffusion would be stable) becomes unstable and gives way to a spontaneous, spatially heterogeneous pattern~\citep[cf.][Chap. 2]{murray2mathbiol}. 
Turing then postulated that a family of growth determinants (broadly speaking, \textit{morphogens}) capable of self-organizing heterogeneously could serve to generate complex forms by stimulating growth in a non-uniform manner. 

Turing's equations and other \textit{reaction-diffusion systems} reveal, in essence, how a simple chemical mechanism based on diffusion---a process normally expected to homogenize concentrations---can drive heterogeneity, offering a window into \textit{emergence}, a key (albeit ambiguously defined) concept in biophysics. 

Phyllotaxis provides a paradigm example of such dynamical self-organization in plants~\citep{godin2020phyllotaxis}. 
 Turing developed an early interest in the topic, which he recognized as a promising application for a morphogen-based model 
~\citep[a work cut short by his death in 1954 and published posthumously as fragmented notes in 1992, 
albeit with relatively marginal impact; cf.][]{swinton2004watching,rueda2014alan}. Several authors have since extended his approach through more or less detailed continuum models~\citep{meinhardt1998models,smith2006inhibition,NEWELL2008421,rueda2018curvature} or through discrete cell-scale descriptions~\citep{jonsson_auxin-driven_2006,smith_plausible_2006,de2006computer,cieslak2015auxin,hartmann2019toward} integrating polar auxin transport by PIN proteins. 
The broad paradigm of spontaneous morphogen organization through inhibitory fields has become increasingly influential in phyllotaxis, with auxin redistribution by PIN transporters now a central focus of research {\citep{traas_phyllotaxis_2013, Vernoux.2021}}.

Yet within the context of morphogenesis, the knowledge of the spatial organization of morphogens does not directly reveal the form of the organism that arises from this organization. 
Turing already had a good grasp of the issue, which he described as \blockquote[]{a problem of formidable mathematical complexity}:
\blockquote[{\citep{turing1952chemical}}]{In determining the changes of state one should take into account: (i) the changes of position [...] as given by Newton's laws of motion; (ii) the stresses as given by the elasticities [...] taking into account the osmotic pressures as given from the chemical data; (iii) the chemical reactions; (iv) the diffusion of the chemical substances [...]}
This problem is inherently mechanical and especially hard; thus, as he recommended, it often necessitates the use of \textit{digital computers}, a path followed with great flair by his successors. 

In their paper \textit{The genetics of geometry},~\cite{coen_genetics_2004} proposed a simple computational paradigm for linking genes to form. In this model, 
the action of genes is to control the local kinematic properties of growth directly. For example, a given gene with a high expression level in a tissue region may result in faster or slower growth. Similarly, the anisotropic expansion of a given region may be prescribed through the action of a \textit{growth polarizer} defining a preferential growth direction locally. 

At the root of this notion is the observation that the development of a general form can be {locally} broken down into a \textit{finite} fundamental developmental `vocabulary' given in terms of kinematic descriptors (\cref{morphology}): growth rate, anisotropy and direction, plus a rotation. 
The idea is then to link gene activity to these kinematic variables through explicit constitutive laws. 
This so-called paradigm of \textit{specified growth} has since become conceptually influential and has been applied to many case studies, informed by experimental genetics and imaging. 

Yet, the link between genes and growth, that is, the causative chain of events connecting gene chemistry to growth mechanics across multiple scales, is not captured in this approach. 
Thus, such \textit{morphogenetic} models may be regarded as phenomenological, in the sense that their focus is on the morphological consequences of a given specified growth field, though without express consideration for its physical causes or feasibility. 

Mathematically, there exists a fundamental difficulty in specifying growth. Indeed, it is well-known that an arbitrary specification of a strain field does not yield a compatible deformation in general (loosely speaking, \textit{the patches do not fit together when rejoined after growth}), unless so-called geometric \textit{compatibility conditions} are met~\citep[cf.][Chap. 2]{barber2002elasticity}. To alleviate this issue, Coen and coworkers introduced the notion of \textit{resultant growth}, referring to the deformation obtained by computationally correcting the specified growth field at each time step to satisfy these constraints, typically via an additional elastic energy minimization step~\citep{kennaway_generation_2011}. Due to the discrepancy between the (reference) specified and resultant growth, this intermediate step leads to the build-up of internal stress, i.e., internal cohesion forces arising from growth incompatibility. However, in most instances considered by authors, this stress is discarded from the computation, so that the resulting configuration at each time step is ultimately stress-free, preventing the accumulation of stress in the tissue as growth simulation proceeds. Then a new specified growth step can be performed, and the process continues iteratively. 
In fact, this growth procedure is framed in algorithmic terms, not fully translating into a mathematical structure, e.g. a system of partial differential equations, amenable to analytical treatment 
(in particular, the existence of a continuous-time limit for the stress-release procedure is unclear). 

In the study of morphogenesis, a tradition in continuum mechanics gave rise to the theory of \textit{morphoelasticity}~\citep[a term coined by Alain Goriely in his 2005 lecture at the \textit{Rencontre du Non-Lin\'eaire} in Paris; cf.][]{goriely2005morpho}. Morphoelasticity is a mechanical theory of growth built upon nonlinear elasticity and plasticity, seeking to formalize the geometry, mechanics, and thermodynamics of a growing body mathematically. This approach offers a natural pathway towards a field theory of plant growth. 

\section{Morphoelasticity: the mechanics of growth}  

The understanding that form arises from forces has led authors to approach growth as a problem of solid mechanics, an effort that entails formulating appropriate balance laws and examining the physical nature of growth. In this context, the emphasis shifts from defining growth as a change in form, towards defining it more fundamentally as a change in \textit{mass}, a perspective which offers a more direct link between physiology and form, and a stronger connection with the open-system thermodynamics of a growing body~\citep{ambrosi2011perspectives,ambrosi2019growth}.

\begin{figure}[ht!]
 \centering
 \includegraphics[width=\linewidth]{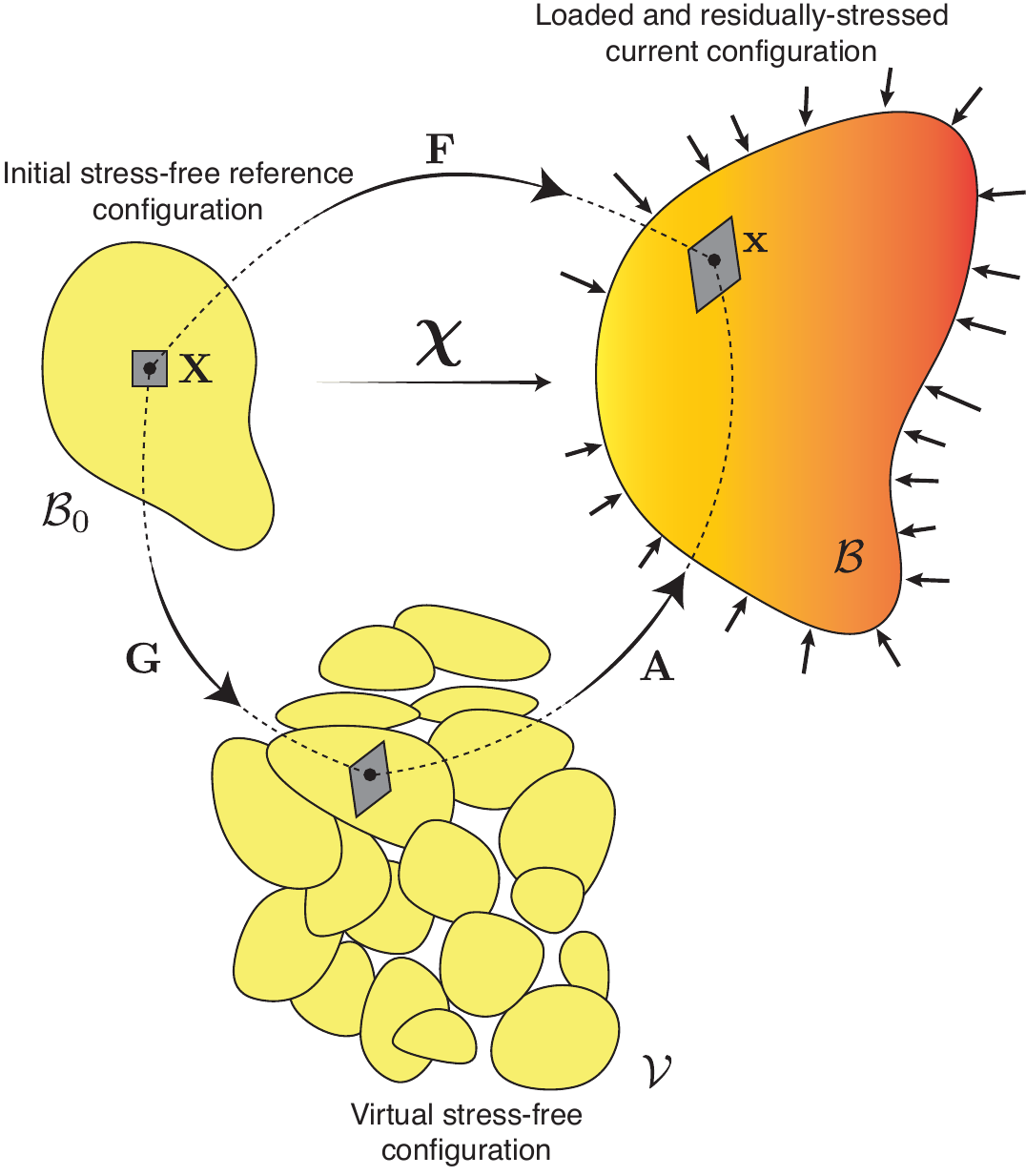}
 \caption{The multiplicative decomposition of morphoelasticity. Starting from a stress-free initial configuration, a local growth deformation $\matr G$ is applied on volume elements, resulting in an incompatible intermediate configuration. A second deformation $\matr A$ ensures compatibility, and results in a stressed configuration that includes residual growth stresses and external loads.}
 \label{fig:morphoelasticity}
 \end{figure}

Translating a local gain in mass into a global deformation has been an important problem in mechanics. This has commonly been achieved through the introduction of the \textit{growth tensor}~\citep{rodriguez1994stress} which quantifies the accumulated change in the resting configuration of a given infinitesimal volume upon local mass accumulation and reorganization. This growth tensor field (henceforth denoted $\matr G$) is not compatible in general (i.e. it does not derive from a deformation map; \cref{morphology}). Therefore, it is combined with an additional \textit{elastic deformation tensor} ($\matr A$). Constitutively, the deformation gradient $\matr F$ at a given point is thus taken to reflect the two composed contributions given by the product
\begin{equation}\label{eqn:multiplicative}
 \matr F 
 = \matr A \matr G.
\end{equation}
This instantaneous relation expresses the conceptual hypothesis of morphoelasticity known as \textit{multiplicative decomposition} (\cref{fig:morphoelasticity}). 
Physically, the growth tensor $\matr G$ captures the slow \textit{anelastic} expansion of the tissue through mass addition and remodelling. 
This component is associated with an intermediate, stress-free configuration, often loosely interpreted as a collection of disjoint and stress-free volume elements 
\cite[cf.][Chap. 12]{Goriely2017}. 
In contrast, the elastic deformation tensor $\matr A$ reflects the rapid elastic deformation of the growing constituents necessary to maintain the integrity of the body. This deformation may be associated with residual mechanical stresses~\citep[details on compatibility are given in][]{jones2012modeling,Goriely2017}.\footnote{We have taken some pedagogical licence in defining incompatibility loosely in terms of \textit{patches not fitting together}. This interpretation does not strictly capture the geometric notion of compatibility. In particular, a necessary condition for \textit{local} compatibility is the vanishing of the curl of $ \matr{G}$~\citep{yavari2013compatibility}, a condition that may be violated in certain stress-free deformations 
\citep{chen2020stress, chen2021physical, dai2022minimizing}.
}

The existence of internal mechanical stress in plants has long been acknowledged~\citep{kutschera1989tissue,peters1996history,kutschera_epidermal-growth-control_2007,LAPOINTE2025102759}. 
The general understanding that tissues may be mechanically stressed by growth itself has generated a wealth of problems, particularly around the exploration of \textit{growth-induced instabilities}---the loss of stability a body experiences at a critical growth threshold leading to a qualitative change in shape, such as buckling~\citep{liang2009shape,dervaux2008morphogenesis,sharon2010mechanics,guo2025midveins,huang2018differential}, wrinkling, creasing~\citep{ben2025wrinkles}, cusping~\citep{zhang2025geometrically}, or tendril perversion~\citep{goriely1998spontaneous}. 
The mathematical study of growth-induced instabilities 
has provided crucial tools to understand the solid mechanics of morphogenesis, and the role of incompatibilities in the emergence of complex forms. 

In their simplest instance, morphoelastic models have focused on a growth field prescribed as a bifurcation parameter governing the onset of instability. In other words, their focus is on the elasticity problems generated by the presence of a growth field, rather than the origin or dynamics of this growth field. In contrast, the problem of \textit{morphodynamics}, covered next, is to model morphogenesis as a \textit{dynamical system}, where growth reflects the evolution of various state variables (e.g. mechanical stress, evolving fields of morphogens), i.e. to formulate a \textit{coupled} theory of growth. 

\section{Morphodynamics: towards a coupled theory}

The general paradigm of morphodynamics is to describe the emergence of form as a coupled dynamical system, with growth coupled with other variables. This perspective builds upon a self-organizing view of developing life where growth emerges as the manifestation of more fundamental physical processes operating within the body. Mathematically, the general problem is then (i) to integrate the physical mechanisms affecting the growth dynamics; and (ii) to formulate growth laws which couple the growth kinetics to other evolving variables of the system.
 
For instance, in plant tropism, a shoot detects an external stimulus (e.g. gravity or light) through specialized cells, which, in response, affect the distribution of auxin in the tissue. Then, auxin stimulates differential growth in the tissue, eliciting curvature and altering the plant's overall posture with respect to the stimulus. This loop can be captured through multiscale modelling and its emergent dynamics can be studied~\citep{chauvet2019revealing,moulton2020multiscale,Oliveri2024active}. In growing tissues, various feedback mechanisms have been postulated between stresses and cell mechanical~\citep{hamant_developmental_2008} and chemical~\citep{heisler2010alignment,nakayama2012mechanical} polarities, and models have been designed to study the behaviour of these feedbacks~\citep{hamant_developmental_2008,alim_regulatory_2012,bozorg_stress_2014,hervieux2016mechanical,oliveri2019regulation,khadka2019feedback,fruleux2019modulation,zhao2020microtubule,ramos2021tissue}. In these scenarios, growth dynamics results from the integration of multiple factors, with their combined effect giving rise to non-trivial emergent properties.

More fundamentally, growth itself is a mechanical phenomenon, and a crucial problem is to couple the rate of tissue expansion to more fundamental physical and mechanical fields~\citep{vandiver2009morpho}. This is the problem of the \textit{growth law}. 
A common perspective has been to view cell wall expansion as a plastic-like yield to pressure forces, combined with remodelling through secretion of new wall material. While the molecular details of this expansion are relatively well described~\citep{cosgrove2005growth,cosgrove2018diffuse}, how exactly this anisotropic yield occurs, and how to model it in multiple dimensions and at the scale of the entire cell or tissue, is rather unclear. While detailed homogenized models for wall expansion in one dimension exist~\citep{dyson_model_2012,smithers2024continuum}, tissue-scale computational models have represented wall expansion in multiple dimensions based on more phenomenological approaches. In particular, linear \textit{strain}-based growth laws allow for easily accounting for the role of cellulose microfibrils in modulating anisotropic growth~\citep{boudon_computational_2015,bozorg_continuous_2016,silveira2025mechanical}. However, biophysically, such phenomenological laws are insufficiently connected to cell wall micromechanics~\citep{cosgrove2025plant}, and, further, are not thermodynamically equivalent to a dissipative plastic yield because they may necessitate an entropy sink to operate
~\citep{oliveri2025field}. Alternatively, authors have adapted the theory of linear plasticity directly to plant walls, e.g. in the context of pollen tube~\citep{dumais2006anisotropic}. Overall, determining an appropriate and well-accepted growth law for cell walls remains an open problem.
 
Building \cor{growth models} 
in a continuum where cells are not explicitly represented is especially difficult. To circumvent this challenge, multicellular computational models have been developed to represent the effect of pressure on individual cells explicitly~\citep[e.g.][]{rudge2005computational,dupuy_system_2008,merks2011virtualleaf,fozard_vertex-element_2013,boudon_computational_2015,cheddadi2019coupling}. 
 These models provide a refined mechanistic view of tissue expansion, enabling a more explicit integration of the wall mechanical properties and cellular topology, and avoiding the need to prescribe growth directly. However, while providing detailed insight, they are inherently computational and lack the generality, minimalism, scalability, and analytic tractability afforded by a continuum mathematical framework.
 
Further, turgor pressure in most of these models is treated as a \textit{non-dynamical} component, either constant in every cell, or enforced as an explicit function of time. Thus, they may be regarded as \textit{specified turgor models}. While this simplification largely facilitates the numerical treatment of these systems, it nonetheless runs counter to general mechanical sense. In mechanics, hydrostatic pressure is a variable associated with a volume conservation relation; thus, it is typically not directly prescribable. 
A critical reevaluation of the role of water in growth and the nature of turgor is (re-)emerging in the community at both experimental and theoretical levels~\citep{cheddadi2019coupling,long2020cellular,dumais2021,ali2023revisiting,zhang2024biomechanics,laplaud2024assessing,alonso2024water,oliveri2025field}. In the next section, we discuss the challenge of modelling turgor pressure and how this discussion can serve as a basis to advance plant modelling towards a \textit{hydromechanical theory of morphogenesis}. 

\section{Multiphysics: towards a hydro-chemo-mechanical theory}

\subsection{The hydromechanical basis of cell expansion\label{hydrodynamical-basis}}

The development of a physically grounded theory of morphogenesis, whether discrete at the cellular scale or continuum-based, requires a focus on the basic cellular physiology of growth.

The understanding that mechanical forces associated with turgor pressure power the expansion of cells traces back to the work of \cite{schwendener1878mechanische} and \cite{sachs1882text} in the 19\textsuperscript{th} century~\citep[cf.][]{hamant_mechanics_2010}. 
This notion, referred to as \textit{turgor-driven growth}, is now relatively well-accepted; yet, its details and interpretation have sometimes generated confusion, and ignited various debates, e.g., around the notion of the \textit{driving force} of growth, between physiologists Hans Burström, and Peter Ray, Paul Green and Robert Cleland, as reflected in correspondences published in \textit{Nature}~\citep{burstrom1971wishful,ray_role_1972}. In his 1971 letter, Burström expresses scepticism regarding the prevailing notion of {turgor-driven growth}, observing that \blockquote[{\citep{burstrom1971wishful}}]{The rigidity of the walls preventing the entry
of water is the cause of the turgor pressure [...] The driving force of any expansion is a difference in water potentials. Expansion is due to water uptake.} 
As a provocative conclusion, he recommends that \blockquote[]{The literature on plant
cell growth would certainly improve if the notion of turgor
expanding the cell was abandoned and replaced by accepted
equations for water balance of fluxes.}
Burström's perspective is partly correct: fundamentally, cells grow by absorbing water, which both \textit{causes} and controls the build-up of turgor pressure, a process governed by mass balance. 
\cor{Indeed}, hydrostatic pressure builds up as an effect of cell walls mechanically resisting water uptake. But without an explicit constitutive assumption for the cell wall---which must expand to allow for water intake---this principle alone is insufficient to complete the picture. 
In their response, Ray et al. write: \blockquote[{\citep{ray_role_1972}}]{Burström fails to come to grips with the principle that irreversible increase in plant cell volume involves simultaneous water uptake (driven by a water potential difference) and cell
wall yielding that depends on turgor stress, and in this sense is ``driven by'' turgor pressure.} 
They conclude---this time advancing a more explicit hierarchy---that
\blockquote[]{Clearly stress relaxation is the primary event in cell enlargement,
whereas water uptake, volume increase and extension
(strain) of the cell wall are secondary.} Here, Ray and colleagues emphasize the \textit{rheological} nature of growth, noting that for cells to expand, their walls must yield to make room for water, a phenomenon indeed caused by turgor. 

At the root of these controversies lies the model of \cite{lockhart1965analysis} proposed a few years earlier, which offers a systematic bridge between wall mechanics, water uptake and turgor. 
Lockhart's seminal approach has seen numerous extensions, notably by \cite{cosgrove1981} and \cite{ortega_augmented_1985}, somewhat in a similar manner. 
The extended model can be summarized as follows. We consider the elongation of a long cylindrical cell of length $\ell$, wall thickness $\delta$, cross-sectional perimeter $\mathcal P$, and cross-sectional area $\mathcal A$ (such that $\delta \ll \mathcal A/\mathcal P \ll \ell $). We introduce the cell volume $\mathcal V$ and cell surface $\mathcal S$.

The expansion rate $\dot \ell $ of the cell can be expressed in terms of the volumetric flux of water across the thin wall, through the \textit{balance of mass equation}~\citep{DAINTY1963279}
\begin{equation}
\dot {\mathcal V} = \frac{k^*\mathcal S }{ \delta}(\pi-p) , \label{eqn:ortega-fluxes-general}
\end{equation}
with $k^*$ the hydraulic conductivity of the cell wall; $\pi$ and $p$ respectively the excess osmotic and hydrostatic pressures relative to the outside. 
The r.h.s. in \eqref{eqn:ortega-fluxes} corresponds to the osmotic influx of water; with the quantity $\psi = p-\pi $ denoting the \textit{water potential} of the cell relative to the outside, measuring the free energy of water 
\citep{niklas2012plant,nobel2020,forterre2022basic}. Using $\mathcal V=\mathcal A\ell $ and $ \mathcal S  \approx \mathcal P \ell $ and rearranging the terms, we obtain 
\begin{equation}
\frac{\dot \ell}{\ell} \approx k   \pp{\pi - p} , \label{eqn:ortega-fluxes}
\end{equation}
with $k\eqdef  k^* \mathcal P /  \mathcal A \delta$ the \textit{effective} hydraulic conductivity of the cell. 

Note that \eqref{eqn:ortega-fluxes} does not define a closed system, as the pressure $p$ is related to the mechanics of the cell wall via the balance of forces between the wall and the water content. 
To allow for water influx, the cell wall must expand, which involves loosening, yielding under tension, and remodelling of the cell walls~\citep{cosgrove2005growth}. This process may be modelled through an effective, Maxwell-type visco-elasto-plastic rheological law for the cell wall,
\begin{equation}
 \frac{\dot \ell }{\ell} = \chi \ramp{\varepsilon - \varepsilon_y} + \dot \varepsilon,\label{eqn:ortega_strain}
\end{equation}
with $\varepsilon$ the cell wall elastic strain; $\varepsilon_y$ a threshold strain above which wall yield and synthesis occur, as captured by the ramp function $\ramp{x}\eqdef\max\pp{x,0}$; and $\chi$ the chemical rate corresponding to wall synthesis. 
In this cylindrical cell considered in quasi-static equilibrium, the pressure $p$ and longitudinal wall stress $\sigma$ are linked through 
\begin{equation}
    \mathcal P\delta \sigma = \mathcal A p. \label{eqn:lockhart-equilibrium}
\end{equation}
On combining this relation with Hooke's law,
\begin{equation}
    \sigma = E^* \varepsilon, \label{eqn:young}
\end{equation}
which links tension to elastic strain via the cell-wall Young's modulus $E^*$, we obtain the pressure-strain relation
$p = E\varepsilon$, with $E\eqdef  E^*  \mathcal P\delta/\mathcal A$ 
measuring the cell's effective elastic stiffness under pressure loads. Thus,
\eqref{eqn:ortega_strain} can be recast as 
\begin{equation}
 \frac{\dot \ell }{\ell} = \phi \ramp{p - y} + \frac{\dot p}{E} ,\label{eqn:ortega}
\end{equation}
with $y \eqdef E\varepsilon_y$ a yield threshold pressure; and $\phi \eqdef \chi/E$ the so-called \textit{extensibility} of the cell. 
Equations \eqref{eqn:ortega,eqn:ortega-fluxes} now form a closed system for $\ell$ and $p$, for which a solution is straightforward to derive. 

In the steady growth regime with constant pressure ($\dot p = 0$), $p$ is fully determined by the three parameters $y$, $\pi$ and $k$ as
\begin{equation}\label{eqn:lockhart-pressure}
 p = \frac{k\pi+\phi y}{k+\phi}, 
\end{equation}
and we obtain \textit{Lockhart's equation} from \eqref{eqn:lockhart-pressure,eqn:ortega}:
\begin{equation}\label{eqn:lockhart-eqn}
 \frac{\dot \ell}{\ell} = \frac{k\phi}{k+\phi} \ramp{\pi - y}.
\end{equation} 
This single equation captures the cell elongation under quasi-static
pressure conditions, expressed in terms of physiological and
rheological parameters. Importantly, \eqref{eqn:lockhart-eqn} applies
only to a \textit{single} elongating cell. This is, for instance, the
case of hair cells of cotton, which can increase their volume by up to
1,000-fold compared to their initial meristematic size
\citep{ruan2001, cosgrove2005growth, hernandez2024}. 

Turgor pressure here results from an equilibrium between mechanical (elastic) and osmotic forces controlled by $\phi$ and $k$ through \eqref{eqn:lockhart-pressure}. By comparing the magnitude of these parameters, we identify two distinct regimes: (i) a wall-limited regime where $\phi \ll k$ and $p \approx \pi$; and (ii) a flux-limited regime with $\phi \gg k$ and $p \approx  y$~\citep{cheddadi2019coupling,dumais2021,ali2023revisiting}. In intermediate situations, we have $y \leq p \leq \pi$ (if $y>\pi$, the cell cannot grow). A common, albeit debated assumption is that, in most scenarios relevant to morphogenesis, the cell operates in the wall-limited regime, so that the approximation
\begin{equation}\label{eqn:approx-pressure}
 p \approx \pi
\end{equation} 
holds. 
In this situation, the cell is close to hydraulic equilibrium and turgor pressure is fully prescribed by the cell chemistry via $\pi$. 
Yet, the relative contributions of water fluxes and wall synthesis to growth control remain unclear, and recent works tend to build a more nuanced picture~\citep{laplaud2024assessing} where hydraulic resistance cannot be neglected. In this scenario, growth, sustained through continuous osmolite supply, maintains cells in a state of hydraulic imbalance with $p < \pi$. In other words, growing cells are out-of-equilibrium systems, in which the chemical energy from the osmolites is dissipated through water transport and wall extension.
Growth and turgor both arise as emergent properties of this process.

We can now revisit Burström's objection: Since turgor is inherently dynamical and lacks a straightforward correlation with growth rate, the notion of turgor-driven growth contributes to an inaccurate picture of pressure viewed as an external force decoupled from the mechanics of the cell and from water transport. 

\subsection{From cell physiology towards a multiphysics theory of tissue morphogenesis\label{multiphysics}}

 To advance our understanding of tissue growth, we must build \textit{multiphysics} theories that combine the diverse physical and mechanical processes that control morphogenesis within a single closed mathematical framework. Specifically, the challenge in plants is to integrate morphogens, growth, mechanics, and hydraulics on the basis of physical principles. To that end, we must first challenge the specified-growth paradigm, which oversimplifies tissue mechanics at a fundamental level.  
 Lockhart's model provides a physiology-based paradigm for plant growth that is now widely accepted in the plant community and has been directly extended to various applications, including growth models at the organismal scale, 
 for example, modelling a fruit as a single Lockhart-type compartment~\citep{fishman_genard_1998}. Such compartment-based models seek to describe water and sugar relations between sinks (growing organs, fruits) and sources (roots, leaves) in complex organs, yet without any explicit dependency on the geometry. \cor{Other authors have attempted to extend Lockhart's model directly to a phenomenological tensorial growth law for a continuum; however, this approach relies on an ad hoc tensorial notion of pressure which does not have a clear mechanical basis~\citep{pietruszka2007anisotropic,lewicka2007anisotropic}.} 

\cor{We stress that} Lockhart's model describes a cylindrical cell expanding longitudinally, for which the stress-pressure relation is independent of the length of the cell, yielding a description of growth kinetics directly in terms of pressure. In particular, the behaviour of this cell is described in terms of effective parameters $k$, $\phi$, $y$, and $E$, which are not \textit{intensive} properties, as they depend on the cross-sectional geometry of the cell. Such a description does not directly apply to other cell shapes where the surface-to-volume ratio varies with growth and subtle geometric effects may influence the stress-pressure relationship as well (this is familiar to anyone who has blown into a spherical rubber balloon, where pressure first increases up to a threshold in radius, and then decreases). Thus, any effort to extend the hydromechanical phenomenology of Lockhart's model should start by separately describing (i)  water transport and gain  (balance of mass), as in 
\eqref{eqn:ortega-fluxes-general}; (ii) mechanical equilibrium (balance of momentum) between cell pressure and wall stress, as in \eqref{eqn:lockhart-equilibrium}; and (iii) cell wall growth and elasticity constitutive relations, as in \eqref{eqn:ortega_strain,eqn:young}.

This principle is crucial for extending the discussion to multicellular tissues, in which additional, spatially-extended collective hydromechanical effects may arise. Lockhart predicted that, in a tissue, \blockquote[\citep{lockhart1965analysis}]{[t]he resistance of the plant to the flow of water from the source of water to the growing tissue will, in general, exert a marked influence on cell elongation}. In contrast to a single cell growing on a hydrated medium, the growth of a region within a tissue may be hindered by hydraulic effects depending on its location within the tissue. 
In such a \textit{poro-morpho-elastic} material, dimensional analysis indicates that pressure should vary over a characteristic hydromechanical length of order $\sim \sqrt{K G / \chi}$ setting the lengthscale of hydraulic interactions~\citep{oliveri2025field}. Here $K$ (with dimensions of \pascal\textsuperscript{-1}.\meter\textsuperscript{2}.\second\textsuperscript{-1}) and $G$ (\pascal) are respectively the bulk water permeability 
 and elastic shear modulus at the tissue scale, 
 and $\chi$ (\second\textsuperscript{-1}) is the chemical rate of cell wall synthesis of \cref{hydrodynamical-basis}. 

Many works---ours included---using Lockhart's rheological equation \eqref{eqn:ortega-fluxes} as a premise, typically proceed with a multicellular model in which a fixed pressure is prescribed in every cell, implemented as a constant akin to other constitutive parameters such as the cell extensibility or Young's modulus. By construction, this simplification precludes hydraulic effects and, in this sense, actually departs from Lockhart's general view. This modelling choice assumes that all cells in the tissue are in the wall-limited regime \eqref{eqn:approx-pressure}, i.e. assuming that the cells grow slowly and close to hydraulic equilibrium. 
While this simplification likely offers a reasonable approximation in many cases, it immediately excludes a broader spectrum of richer regimes.

The hydrostatic pressure of a cell is linked to the mechanics of cell walls, their geometry, their location within the mechanical context of the tissue, and water fluxes governed by balance relations. 
Any change in these properties, e.g. through modifications of 
cell osmolarity, wall extensibility, permeability, or elastic moduli, may, in principle, generate a change in cell pressure. 
These consequences of Lockhartian hydrostatics have the potential to influence the interpretation (or misinterpretation) of experiments and simulations significantly, underscoring the need for a thorough rethinking and rigorous treatment of pressure in both theoretical and experimental works. 

The actual extent of water effects in morphogenesis is still poorly characterized, and constitutes an active subject of research. 
Thus, to deepen our understanding of tissues, a prudent approach is to dispense with the costly assumption of prescribed pressure, with the idea that a sound, parsimonious physical theory should remain agnostic about the ongoing tensions between concurrent biological hypotheses. 
In recent years, we, with other colleagues, have sought to revive the debate and revisit the dynamical properties of plant matter through the lens of hydromechanical principles. 

\cite{cheddadi2019coupling} proposed a vertex-based
cellular-level model
including hydraulic fluxes between cells,
growth and elasticity, thereby extending Lockhart's approach
directly. In this model, wall tensions and pressure are coupled through mechanical equilibrium, and the growth of the cell walls results from wall tension. In a tissue, this tension reflects both the cell hydrostatic pressure and the mechanical influences of neighbouring cells~\citep{boudon_computational_2015}. In all cases, any change of volume has to be accommodated by water flux,
as reflected by \eqref{eqn:ortega-fluxes}, where pressure appears as a contribution to the cell water potential.
Hence, pressure adapts to both mechanical and hydraulic constraints, and it is indeed a dynamical variable. As in Lockhart's model, turgor pressure is lower than the osmotic pressure for a growing cell: growth brings cells out of equilibrium, so that  turgor pressure can become heterogeneous
between cells with finite water conductivity.

A striking example of such heterogeneous turgor distribution appears in tissues with heterogeneous cell topology, that is, in irregular tissues where growing cells have a variable number of neighbours~\citep{long2020cellular}. In principle, all other things being equal,   cell geometry and mechanical balance are expected to result in slower growth of cells with fewer neighbours~\citep{ali2023revisiting}.  Furthermore, these cells are expected to have higher turgor pressure, even though they are hydraulically connected to their neighbours. This phenomenon is well known in foams~\citep{cantat2013foams}, and was
also indirectly observed in oryzalin-treated meristems where cells with fewer neighbours have a convex shape, bulging outward into their neighbours, potentially indicative of higher turgor~\citep{corson_turning_2009}. 
\cite{long2020cellular} confirmed this interpretation with atomic-force-microscopy-based measurements of untreated and oryzalin-treated meristems, giving estimates for individual cell pressures.
In foams, the pressure difference between bubbles is associated with gas exchanges, and smaller bubbles (with generally fewer neighbours and higher pressure) tend to shrink and eventually
disappear, contributing to the growth of their larger neighbours, through the so-called von Neumann-Mullins coarsening~\citep{cantat2013foams}. Plant cells differ from bubbles in that 
they possess an osmotic potential that prevents them from
shrinking in general. Yet, one may anticipate that smaller cells would be less
able to grow than their larger neighbours.
This was observed in oryzalin-treated
meristems~\citep{long2020cellular}, but surprisingly not 
in untreated ones. Indeed, smaller cells also benefit from a higher surface-to-volume ratio. In the untreated case, this hydraulic advantage may suffice to
overcome the mechanical disadvantage arising from topology. Oryzalin
could alter this balance. Interestingly, such a shift between these two regimes was
also observed by \cite{tsugawa2017clones} in growing \textit{Arabidopsis thaliana} sepals, with smaller cells growing faster at early stages and the opposite
later on. Whether this shift is regulated by a global change in the hydraulic and/or mechanical properties of the sepal remains unknown. However, these observations combined with modelling, provide a clear example of how moving beyond the prescribed-turgor paradigm ($p = \pi$) opens a window on new emergent properties and enriches the theoretical vocabulary available to model the regulation and patterning of growth, and thus deepens our understanding of these processes, potentially opening new avenues of experimental investigations.

As previously discussed, pressure adjusts to mechanical constraints, and cells with lower wall tension generally have lower turgor pressure. This effect can appear in primordia at the apical meristem: cells in this region have a much
higher growth rate, resulting in the bulging of the primordia~\citep {kwiatkowska2003growth}, likely caused by cell wall loosening in these cells~\citep{kierzkowski_elastic_2012,sassi_auxin-mediated_2014}. Then, lower turgor in these cells results in lower water potential, giving them an additional hydraulic
advantage. This effect was analysed by \cite{cheddadi2019coupling}
who showed that growing primordia act as water sinks that
pump water from their neighbourhood in virtue of their lower turgor, in a way reminiscent of the growth-induced water potential concept
developed by John Boyer and coworkers~\citep{molz1978growth}. \cite{cheddadi2019coupling} showed that a situation of scarce water resources further amplifies competition between
neighbouring cells. When cell-cell hydraulic connectivity is important, 
this flux-based lateral inhibition can create a growth rate heterogeneity large enough to generate the sharp change of tissue curvature at the primordium
boundary seen in meristems. Strikingly, this effect can even result in boundary cells shrinking. Conversely, if the cells have unlimited access to water or if water exchanges between cells are weak, no hydraulic competition occurs, and the simulated curvature change at the boundary appears shallower.

This consequence of hydraulics was recently confirmed by
experimental observations of shrinking
cells in the apical meristem at the boundary of growing primordia, which was also shown
to contribute to defining the cellular identity of the
boundary domain~\citep{alonso2024water}.
Here again, the coupling between wall growth and mechanics and water transport affects the growth dynamics and the resulting form deeply. These results collectively reveal
new potential regulatory mechanisms for morphogenetic patterns.

Recently, this modelling work was extended to a \textit{continuum} theory~\citep{oliveri2025field}, combining morphoelasticity with poroelasticity, a theory of fluid-saturated solids~\citep{forterre2022basic}. This work synthesizes previous exploration by \cite{philip1958propagation,molz1974water,molz1975dynamics,molz1978growth,silk1980growth,plant1982continuum,passioura2003tissue,wiegers2009modeling} and establishes a field theory of growth in plants. In this model, the fluid and solid phases of the tissue are both modelled via explicit balance relations and constitutive assumptions; thus, the movement of water, in relation to the expansion of the tissue, can be included, enabling unbalanced water potentials and heterogeneous pressures. Integrating the phenomenology of Lockhart's model within a continuum framework enables a mechanistic description of tissue growth dynamics. This approach relies on explicit, coupled laws that link cell expansion to stresses, strains, and pressure, thereby moving beyond models that directly prescribe the growth tensor or rely on computational cell-based formulations. 

This continuum formulation thus provides an analytic view on the macroscopic laws governing tissue expansion, which feature explicit biophysical parameters, ultimately controlled by genes. For instance, this model yields a complete analytic description of the phenomenon of water competition generated by a heterogeneous stiffness, for which a closed-form asymptotic solution for the pressure and growth profiles can be derived. In contrast to a cell-based computational description, this approach relies on well-established tools of nonlinear solid mechanics in three dimensions, allowing us to describe the finite deformations of entire organs rigorously while freeing us from the constraints of cell-scale modelling. Based on these concepts, \cite{oliveri2025field} revisited the classic problem of tissue tension in shoots \citep{peters1996history} through the lens of hydraulic principles \citep[following an earlier idea by][]{passioura2003tissue}. In contrast to specified-growth models, which are essentially timescale-free, this framework portrays shoot expansion as a dynamic process, in which geometry and kinematics are constrained by hydraulics, defining the shoot's attainable space of forms.

\section{Concluding remarks and perspectives}

In retrospect, the tensions surrounding the concept of the driving force reflect a definitional---rather than phenomenological---debate: \blockquote[\citep{MONEY1997173}]{the difference of opinion hinged on the definition of ``driving force'' rather than any disagreement about the events that occur during expansion}. In plant active matter, the driving force of growth---if we are to hold on to that notion---lies in the metabolic activity of the cells, which work to maintain the chemical potential needed to power their osmotic potential and, consequently, their growth. This energy comes from the environment through photosynthesis.  
Turgor and growth reflect a complex equilibrium between various chemical and physical processes, as their values emerge from water and solute fluxes, mechanical equilibrium, and constitutive laws for cell wall elasticity and anelasticity. 
From this perspective, plant living matter may be viewed macroscopically as a poro-morpho-elastic material, introducing a characteristic length scale $\sqrt{K G / \chi}$, reflecting the coupling of hydraulics (permeability $K$), mechanics (elastic modulus $G$), and wall synthesis (chemical rate $\chi$) defining the typical length of hydraulic interactions. This length may, in principle, govern the water-based morphogenetic patterns emerging \cor{through} amplification of growth heterogeneities by water fluxes, as proposed in the shoot apex~\citep{alonso2024water}.

Plants possess all the attributes of active matter: they are open and out-of-equilibrium systems that support non-local emergent behaviours. 
Yet in contrast to animals and their swirling embryos, the active properties of plants remain subtle and reserved, quietly concealed behind their cellulose curtains. To engage with these properties, we must move beyond linear and reductive conceptions of growth mechanics. The general paradigm and concepts of active matter physics enable a conceptual leap towards a more systematic understanding of plant growth. 
In particular, the theoretical approaches emphasized in \cref{multiphysics} highlight the inherently out-of-equilibrium nature of growth in plants, where fluxes of mass and energy remain unbalanced. This imbalance is the condition of plants: \textit{in a perpetual state of growth, a plant continually evades balance.} By incorporating such active properties explicitly, these approaches reveal emergent spatiotemporal couplings that arise directly from hydromechanical principles, e.g., water-competition phenomena between different regions of the tissue. Taken together, they offer a more nuanced picture of the physical nature of plants.
 
The main goal of this perspective paper, overall, was (i) to critically reassess key assumptions of plant modelling, especially the specified-growth and prescribed-turgor paradigms, in light of physical principles and recent experiments, and (ii) propose conceptual tools for developing a theory of \textit{emergence} in plant forms. Important challenges and perspectives towards such a theory include:
\begin{enumerate}
 \item Establishing a growth law based on well-accepted physical processes and first principles, and reproducing a range of experiments. This law should also account for the cellular properties of the tissues systematically; here, multiscale approaches~\citep[e.g.][]{ghysels2010multiscale,boudaoud2023multiscale} will be valuable.
 \item Revisiting the relationship between water and growth, experimentally and theoretically. Understanding the role of water transport in morphogenesis, its pathways, and regulations.
 \item Developing a complete description of the thermodynamics of growth and water dissipation within a tissue.
 \item Extending the description of `naked' plant matter to a theory of \textit{smart active matter}~\citep{levine2023physics}, including self-regulation, viz., mechanical feedbacks, complex hormonal interactions, ion transport, and gene regulation. 
\end{enumerate}

One approach emphasized here involves viewing the tissue as a continuum, which yields simplified and mathematically tractable representations focusing on bulk properties rather than cytohistological details. This approach participates in an \textit{organismal} view of plant multicellularity, 
in which the organism develops independently of cellular structure, and, conversely, cell proliferation is relatively separate from---or even subordinate to---overall growth~\citep{kaplan1991relationship}. 
This perspective supports a cell-free, solid-mechanics view of plant tissues and argues for shifting attention from individual cells seen as independent agents of form---a view long central to gene-focused analyses of morphogenesis---to the global mechanical properties of tissues: \blockquote[\citep{sinnott1939cell}]{The development of an organ proceeds with little relation to the manner in which it is cut up into organized cellular units [...] 
Organization at one level seems independent of that at another.} 

The general effort to rationalize the mechanisms of form mathematically is
crucial for achieving a quantitative leap in our understanding of plant development, to build a unified and well-accepted theory of morphogenesis, and for moving beyond the reliance on ad hoc simulations. Overall, our goal is to build a clearer picture of the constraints that act on plant growth by defining the space of forms physically achievable 
through gene control systematically.

\section*{Acknowledgments} 

\paragraph{Competing interests.} The authors declare no competing interests exist.

\paragraph{Authors contributions.} All authors contributed equally.

\paragraph{Funding statement.} The authors acknowledge support from the \textit{Institut rhônalpin des systèmes complexes} (IXXI). I.C. and C.G. acknowledge support from the French National Research Agency (ANR) under Research Grant \href{https://anr.fr/Project-ANR-20-CE13-0022}{ANR-20-CE13-0022-03} \textit{HydroField}.
 

\begin{thebibliography}{159}
\expandafter\ifx\csname natexlab\endcsname\relax\def\natexlab#1{#1}\fi
\providecommand{\url}[1]{\texttt{#1}}
\providecommand{\href}[2]{#2}
\providecommand{\path}[1]{#1}
\providecommand{\DOIprefix}{doi:}
\providecommand{\ArXivprefix}{arXiv:}
\providecommand{\URLprefix}{URL: }
\providecommand{\Pubmedprefix}{pmid:}
\providecommand{\doi}[1]{\href{http://dx.doi.org/#1}{\path{#1}}}
\providecommand{\Pubmed}[1]{\href{pmid:#1}{\path{#1}}}
\providecommand{\bibinfo}[2]{#2}
\ifx\xfnm\relax \def\xfnm[#1]{\unskip,\space#1}\fi
\bibitem[{Ali et~al.(2023)Ali, Cheddadi, Landrein and Long}]{ali2023revisiting}
\bibinfo{author}{Ali, O.}, \bibinfo{author}{Cheddadi, I.}, \bibinfo{author}{Landrein, B.}, \bibinfo{author}{Long, Y.}, \bibinfo{year}{2023}.
\newblock \bibinfo{title}{Revisiting the relationship between turgor pressure and plant cell growth}.
\newblock \bibinfo{journal}{New Phytologist} \bibinfo{volume}{238}, \bibinfo{pages}{62--69}.
\newblock \URLprefix \url{https://nph.onlinelibrary.wiley.com/doi/10.1111/nph.18683}, \DOIprefix\doi{10.1111/nph.18683}.
\bibitem[{Ali et~al.(2014)Ali, Mirabet, Godin and Traas}]{ali2014physical}
\bibinfo{author}{Ali, O.}, \bibinfo{author}{Mirabet, V.}, \bibinfo{author}{Godin, C.}, \bibinfo{author}{Traas, J.}, \bibinfo{year}{2014}.
\newblock \bibinfo{title}{Physical models of plant development}.
\newblock \bibinfo{journal}{Annual Review of Cell and Developmental Biology} \bibinfo{volume}{30}, \bibinfo{pages}{59--78}.
\newblock \URLprefix \url{https://www.annualreviews.org/content/journals/10.1146/annurev-cellbio-101512-122410}, \DOIprefix\doi{10.1146/annurev-cellbio-101512-122410}.
\bibitem[{Alim et~al.(2016)Alim, Armon, Shraiman and Boudaoud}]{alim2016leaf}
\bibinfo{author}{Alim, K.}, \bibinfo{author}{Armon, S.}, \bibinfo{author}{Shraiman, B.I.}, \bibinfo{author}{Boudaoud, A.}, \bibinfo{year}{2016}.
\newblock \bibinfo{title}{Leaf growth is conformal}.
\newblock \bibinfo{journal}{Physical biology} \bibinfo{volume}{13}, \bibinfo{pages}{05LT01}.
\newblock \URLprefix \url{https://iopscience.iop.org/article/10.1088/1478-3975/13/5/05LT01/meta}, \DOIprefix\doi{10.1088/1478-3975/13/5/05LT01}.
\bibitem[{Alim et~al.(2012)Alim, Hamant and Boudaoud}]{alim_regulatory_2012}
\bibinfo{author}{Alim, K.}, \bibinfo{author}{Hamant, O.}, \bibinfo{author}{Boudaoud, A.}, \bibinfo{year}{2012}.
\newblock \bibinfo{title}{Regulatory role of cell division rules on tissue growth heterogeneity}.
\newblock \bibinfo{journal}{Frontiers in Plant Science} \bibinfo{volume}{3}, \bibinfo{pages}{1--13}.
\newblock \URLprefix \url{https://www.frontiersin.org/journals/plant-science/articles/10.3389/fpls.2012.00174/full}, \DOIprefix\doi{10.3389/fpls.2012.00174}.
\bibitem[{Alonso-Serra et~al.(2024)Alonso-Serra, Cheddadi, Kiss, Cerutti, Lang, Dieudonn{\'e}, Lionnet, Godin and Hamant}]{alonso2024water}
\bibinfo{author}{Alonso-Serra, J.}, \bibinfo{author}{Cheddadi, I.}, \bibinfo{author}{Kiss, A.}, \bibinfo{author}{Cerutti, G.}, \bibinfo{author}{Lang, M.}, \bibinfo{author}{Dieudonn{\'e}, S.}, \bibinfo{author}{Lionnet, C.}, \bibinfo{author}{Godin, C.}, \bibinfo{author}{Hamant, O.}, \bibinfo{year}{2024}.
\newblock \bibinfo{title}{Water fluxes pattern growth and identity in shoot meristems}.
\newblock \bibinfo{journal}{Nature Communications} \bibinfo{volume}{15}, \bibinfo{pages}{1--14}.
\newblock \URLprefix \url{https://www.nature.com/articles/s41467-024-51099-x}, \DOIprefix\doi{10.1038/s41467-024-51099-x}.
\bibitem[{Ambrosi et~al.(2011)Ambrosi, Ateshian, Arruda, Cowin, Dumais, Goriely, Holzapfel, Humphrey, Kemkemer, Kuhl et~al.}]{ambrosi2011perspectives}
\bibinfo{author}{Ambrosi, D.}, \bibinfo{author}{Ateshian, G.A.}, \bibinfo{author}{Arruda, E.M.}, \bibinfo{author}{Cowin, S.}, \bibinfo{author}{Dumais, J.}, \bibinfo{author}{Goriely, A.}, \bibinfo{author}{Holzapfel, G.A.}, \bibinfo{author}{Humphrey, J.D.}, \bibinfo{author}{Kemkemer, R.}, \bibinfo{author}{Kuhl, E.}, et~al., \bibinfo{year}{2011}.
\newblock \bibinfo{title}{Perspectives on biological growth and remodeling}.
\newblock \bibinfo{journal}{Journal of the Mechanics and Physics of Solids} \bibinfo{volume}{59}, \bibinfo{pages}{863--883}.
\newblock \URLprefix \url{https://www.sciencedirect.com/science/article/abs/pii/S0022509610002516}, \DOIprefix\doi{10.1016/j.jmps.2010.12.011}.
\bibitem[{Ambrosi et~al.(2019)Ambrosi, Ben~Amar, Cyron, DeSimone, Goriely, Humphrey and Kuhl}]{ambrosi2019growth}
\bibinfo{author}{Ambrosi, D.}, \bibinfo{author}{Ben~Amar, M.}, \bibinfo{author}{Cyron, C.J.}, \bibinfo{author}{DeSimone, A.}, \bibinfo{author}{Goriely, A.}, \bibinfo{author}{Humphrey, J.D.}, \bibinfo{author}{Kuhl, E.}, \bibinfo{year}{2019}.
\newblock \bibinfo{title}{Growth and remodelling of living tissues: perspectives, challenges and opportunities}.
\newblock \bibinfo{journal}{Journal of the Royal Society Interface} \bibinfo{volume}{16}, \bibinfo{pages}{20190233}.
\newblock \URLprefix \url{https://royalsocietypublishing.org/doi/10.1098/rsif.2019.0233}, \DOIprefix\doi{10.1098/rsif.2019.0233}.
\bibitem[{Arteca(1996)}]{arteca1996plant}
\bibinfo{author}{Arteca, R.N.}, \bibinfo{year}{1996}.
\newblock \bibinfo{title}{Plant growth substances: principles and applications}.
\newblock \bibinfo{publisher}{Springer}, \bibinfo{address}{New York}.
\newblock \URLprefix \url{https://link.springer.com/book/10.1007/978-1-4757-2451-6}, \DOIprefix\doi{10.1007/978-1-4757-2451-6}.
\bibitem[{Avery(1933)}]{avery1933structure}
\bibinfo{author}{Avery, G.S.}, \bibinfo{year}{1933}.
\newblock \bibinfo{title}{Structure and development of the tobacco leaf}.
\newblock \bibinfo{journal}{American Journal of Botany} \bibinfo{volume}{20}, \bibinfo{pages}{565--592}.
\newblock \URLprefix \url{https://bsapubs.onlinelibrary.wiley.com/doi/10.1002/j.1537-2197.1933.tb08913.x}, \DOIprefix\doi{10.2307/2436259}.
\bibitem[{Band et~al.(2014)Band, Wells, Fozard, Ghetiu, French, Pound, Wilson, YU, Li, Hijazi, Oh, Pearce, Perez-Amador, Yun, Kramer, Alonso, Godin, Vernoux, Hodgman, Pridmore, Swarup, King and Bennett}]{Band.2014}
\bibinfo{author}{Band, L.R.}, \bibinfo{author}{Wells, D.M.}, \bibinfo{author}{Fozard, J.A.}, \bibinfo{author}{Ghetiu, T.}, \bibinfo{author}{French, A.P.}, \bibinfo{author}{Pound, M.P.}, \bibinfo{author}{Wilson, M.H.}, \bibinfo{author}{YU, L.}, \bibinfo{author}{Li, W.}, \bibinfo{author}{Hijazi, H.I.}, \bibinfo{author}{Oh, J.}, \bibinfo{author}{Pearce, S.P.}, \bibinfo{author}{Perez-Amador, M.A.}, \bibinfo{author}{Yun, J.}, \bibinfo{author}{Kramer, E.}, \bibinfo{author}{Alonso, J.M.}, \bibinfo{author}{Godin, C.}, \bibinfo{author}{Vernoux, T.}, \bibinfo{author}{Hodgman, T.C.}, \bibinfo{author}{Pridmore, T.P.}, \bibinfo{author}{Swarup, R.}, \bibinfo{author}{King, J.R.}, \bibinfo{author}{Bennett, M.J.}, \bibinfo{year}{2014}.
\newblock \bibinfo{title}{{Systems Analysis of Auxin Transport in the Arabidopsis Root Apex}}.
\newblock \bibinfo{journal}{The Plant cell} \URLprefix \url{http://www.plantcell.org/cgi/doi/10.1105/tpc.113.119495}, \DOIprefix\doi{10.1105/tpc.113.119495}.
\bibitem[{Barber(2002)}]{barber2002elasticity}
\bibinfo{author}{Barber, J.R.}, \bibinfo{year}{2002}.
\newblock \bibinfo{title}{Elasticity}. volume \bibinfo{volume}{172} of \textit{\bibinfo{series}{Solid Mechanics and Its Applications}}.
\newblock \bibinfo{edition}{4} ed., \bibinfo{publisher}{Springer Nature Switzerland}, \bibinfo{address}{Cham}.
\newblock \URLprefix \url{https://link.springer.com/book/10.1007/978-3-031-15214-6}, \DOIprefix\doi{10.1007/978-3-031-15214-6}.
\bibitem[{{Barbier de Reuille} et~al.(2006){Barbier de Reuille}, Bohn-Courseau, Ljung, Morin, Carraro, Godin and Traas}]{de2006computer}
\bibinfo{author}{{Barbier de Reuille}, P.}, \bibinfo{author}{Bohn-Courseau, I.}, \bibinfo{author}{Ljung, K.}, \bibinfo{author}{Morin, H.}, \bibinfo{author}{Carraro, N.}, \bibinfo{author}{Godin, C.}, \bibinfo{author}{Traas, J.}, \bibinfo{year}{2006}.
\newblock \bibinfo{title}{{Computer simulations reveal properties of the cell-cell signaling network at the shoot apex in Arabidopsis}}.
\newblock \bibinfo{journal}{Proceedings of the National Academy of Sciences} \bibinfo{volume}{103}, \bibinfo{pages}{1627--1632}.
\newblock \URLprefix \url{https://www.pnas.org/doi/abs/10.1073/pnas.0510130103}, \DOIprefix\doi{10.1073/pnas.0510130103}.
\bibitem[{{Barbier de Reuille} et~al.(2015){Barbier de Reuille}, Routier-Kierzkowska, Kierzkowski, Bassel, Sch{\"u}pbach, Tauriello, Bajpai, Strauss, Weber, Kiss et~al.}]{de2015morphographx}
\bibinfo{author}{{Barbier de Reuille}, P.}, \bibinfo{author}{Routier-Kierzkowska, A.L.}, \bibinfo{author}{Kierzkowski, D.}, \bibinfo{author}{Bassel, G.W.}, \bibinfo{author}{Sch{\"u}pbach, T.}, \bibinfo{author}{Tauriello, G.}, \bibinfo{author}{Bajpai, N.}, \bibinfo{author}{Strauss, S.}, \bibinfo{author}{Weber, A.}, \bibinfo{author}{Kiss, A.}, et~al., \bibinfo{year}{2015}.
\newblock \bibinfo{title}{{MorphoGraphX: a platform for quantifying morphogenesis in 4D}}.
\newblock \bibinfo{journal}{eLife} \bibinfo{volume}{4}, \bibinfo{pages}{e05864}.
\newblock \URLprefix \url{https://elifesciences.org/articles/5864}, \DOIprefix\doi{10.7554/eLife.05864}.
\bibitem[{Ben~Amar(2025)}]{ben2025wrinkles}
\bibinfo{author}{Ben~Amar, M.}, \bibinfo{year}{2025}.
\newblock \bibinfo{title}{Wrinkles, creases, and cusps in growing soft matter}.
\newblock \bibinfo{journal}{Reviews of Modern Physics} \bibinfo{volume}{97}, \bibinfo{pages}{015004}.
\newblock \URLprefix \url{https://journals.aps.org/rmp/abstract/10.1103/RevModPhys.97.015004}, \DOIprefix\doi{10.1103/RevModPhys.97.015004}.
\bibitem[{Bilsborough et~al.(2011)Bilsborough, Runions, Barkoulas, Jenkins, Hasson, Galinha, Laufs, Hay, Prusinkiewicz and Tsiantis}]{bilsborough2011model}
\bibinfo{author}{Bilsborough, G.D.}, \bibinfo{author}{Runions, A.}, \bibinfo{author}{Barkoulas, M.}, \bibinfo{author}{Jenkins, H.W.}, \bibinfo{author}{Hasson, A.}, \bibinfo{author}{Galinha, C.}, \bibinfo{author}{Laufs, P.}, \bibinfo{author}{Hay, A.}, \bibinfo{author}{Prusinkiewicz, P.}, \bibinfo{author}{Tsiantis, M.}, \bibinfo{year}{2011}.
\newblock \bibinfo{title}{Model for the regulation of {Arabidopsis} thaliana leaf margin development}.
\newblock \bibinfo{journal}{Proceedings of the National Academy of Sciences} \bibinfo{volume}{108}, \bibinfo{pages}{3424--3429}.
\newblock \URLprefix \url{https://www.pnas.org/doi/10.1073/pnas.1015162108}, \DOIprefix\doi{10.1073/pnas.1015162108}.
\bibitem[{Boudaoud et~al.(2023)Boudaoud, Kiss and Ptashnyk}]{boudaoud2023multiscale}
\bibinfo{author}{Boudaoud, A.}, \bibinfo{author}{Kiss, A.}, \bibinfo{author}{Ptashnyk, M.}, \bibinfo{year}{2023}.
\newblock \bibinfo{title}{Multiscale modeling and analysis of growth of plant tissues}.
\newblock \bibinfo{journal}{SIAM Journal on Applied Mathematics} \bibinfo{volume}{83}, \bibinfo{pages}{2354--2389}.
\newblock \URLprefix \url{https://epubs.siam.org/doi/10.1137/23M1553315}, \DOIprefix\doi{10.1137/23M1553315}.
\bibitem[{Boudon et~al.(2015)Boudon, Chopard, Ali, Gilles, Hamant, Boudaoud, Traas and Godin}]{boudon_computational_2015}
\bibinfo{author}{Boudon, F.}, \bibinfo{author}{Chopard, J.}, \bibinfo{author}{Ali, O.}, \bibinfo{author}{Gilles, B.}, \bibinfo{author}{Hamant, O.}, \bibinfo{author}{Boudaoud, A.}, \bibinfo{author}{Traas, J.}, \bibinfo{author}{Godin, C.}, \bibinfo{year}{2015}.
\newblock \bibinfo{title}{A {Computational} {Framework} for 3{D} {Mechanical} {Modeling} of {Plant} {Morphogenesis} with {Cellular} {Resolution}}.
\newblock \bibinfo{journal}{PLOS Comput Biol} \bibinfo{volume}{11}, \bibinfo{pages}{e1003950}.
\newblock \URLprefix \url{http://journals.plos.org/ploscompbiol/article?id=10.1371/journal.pcbi.1003950}, \DOIprefix\doi{10.1371/journal.pcbi.1003950}.
\bibitem[{Bozorg et~al.(2014)Bozorg, Krupinski and J\"{o}nsson}]{bozorg_stress_2014}
\bibinfo{author}{Bozorg, B.}, \bibinfo{author}{Krupinski, P.}, \bibinfo{author}{J\"{o}nsson, H.}, \bibinfo{year}{2014}.
\newblock \bibinfo{title}{Stress and {Strain} {Provide} {Positional} and {Directional} {Cues} in {Development}}.
\newblock \bibinfo{journal}{PLOS Comput Biol} \bibinfo{volume}{10}, \bibinfo{pages}{e1003410}.
\newblock \URLprefix \url{http://journals.plos.org/ploscompbiol/article?id=10.1371/journal.pcbi.1003410}, \DOIprefix\doi{10.1371/journal.pcbi.1003410}.
\bibitem[{Bozorg et~al.(2016)Bozorg, Krupinski and J\"onsson}]{bozorg_continuous_2016}
\bibinfo{author}{Bozorg, B.}, \bibinfo{author}{Krupinski, P.}, \bibinfo{author}{J\"onsson, H.}, \bibinfo{year}{2016}.
\newblock \bibinfo{title}{A continuous growth model for plant tissue}.
\newblock \bibinfo{journal}{Physical Biology} \bibinfo{volume}{13}, \bibinfo{pages}{065002}.
\newblock \URLprefix \url{http://stacks.iop.org/1478-3975/13/i=6/a=065002}, \DOIprefix\doi{10.1088/1478-3975/13/6/065002}.
\bibitem[{Burstr{\"o}m(1971)}]{burstrom1971wishful}
\bibinfo{author}{Burstr{\"o}m, H.G.}, \bibinfo{year}{1971}.
\newblock \bibinfo{title}{Wishful thinking of turgor}.
\newblock \bibinfo{journal}{Nature} \bibinfo{volume}{234}, \bibinfo{pages}{488--488}.
\newblock \URLprefix \url{https://www.nature.com/articles/234488a0}, \DOIprefix\doi{10.1038/234488a0}.
\bibitem[{Cantat et~al.(2013)Cantat, Cohen-Addad, Elias, Graner, H{\"o}hler, Pitois, Rouyer and Saint-Jalmes}]{cantat2013foams}
\bibinfo{author}{Cantat, I.}, \bibinfo{author}{Cohen-Addad, S.}, \bibinfo{author}{Elias, F.}, \bibinfo{author}{Graner, F.}, \bibinfo{author}{H{\"o}hler, R.}, \bibinfo{author}{Pitois, O.}, \bibinfo{author}{Rouyer, F.}, \bibinfo{author}{Saint-Jalmes, A.}, \bibinfo{year}{2013}.
\newblock \bibinfo{title}{Foams: structure and dynamics}.
\newblock \bibinfo{publisher}{Oxford University Press}, \bibinfo{address}{Oxford}.
\newblock \URLprefix \url{https://academic.oup.com/book/9438}, \DOIprefix\doi{10.1093/acprof:oso/9780199662890.001.0001}.
\bibitem[{Chauvet et~al.(2019)Chauvet, Moulia, Legu{\'e}, Forterre and Pouliquen}]{chauvet2019revealing}
\bibinfo{author}{Chauvet, H.}, \bibinfo{author}{Moulia, B.}, \bibinfo{author}{Legu{\'e}, V.}, \bibinfo{author}{Forterre, Y.}, \bibinfo{author}{Pouliquen, O.}, \bibinfo{year}{2019}.
\newblock \bibinfo{title}{Revealing the hierarchy of processes and time-scales that control the tropic response of shoots to gravi-stimulations}.
\newblock \bibinfo{journal}{Journal of Experimental Botany} \bibinfo{volume}{70}, \bibinfo{pages}{1955--1967}.
\newblock \URLprefix \url{https://academic.oup.com/jxb/article/70/6/1955/5420653}, \DOIprefix\doi{10.1093/jxb/erz027}.
\bibitem[{Cheddadi et~al.(2019)Cheddadi, G{\'e}nard, Bertin and Godin}]{cheddadi2019coupling}
\bibinfo{author}{Cheddadi, I.}, \bibinfo{author}{G{\'e}nard, M.}, \bibinfo{author}{Bertin, N.}, \bibinfo{author}{Godin, C.}, \bibinfo{year}{2019}.
\newblock \bibinfo{title}{Coupling water fluxes with cell wall mechanics in a multicellular model of plant development}.
\newblock \bibinfo{journal}{PLoS computational biology} \bibinfo{volume}{15}, \bibinfo{pages}{e1007121}.
\newblock \URLprefix \url{https://journals.plos.org/ploscompbiol/article?id=10.1371/journal.pcbi.1007121}, \DOIprefix\doi{10.1371/journal.pcbi.1007121}.
\bibitem[{Chen et~al.(2021)Chen, Ciarletta and Dai}]{chen2021physical}
\bibinfo{author}{Chen, X.}, \bibinfo{author}{Ciarletta, P.}, \bibinfo{author}{Dai, H.H.}, \bibinfo{year}{2021}.
\newblock \bibinfo{title}{Physical principles of morphogenesis in mushrooms}.
\newblock \bibinfo{journal}{Physical Review E} \bibinfo{volume}{103}, \bibinfo{pages}{022412}.
\newblock \URLprefix \url{https://journals.aps.org/pre/abstract/10.1103/PhysRevE.103.022412}, \DOIprefix\doi{10.1103/PhysRevE.103.022412}.
\bibitem[{Chen and Dai(2020)}]{chen2020stress}
\bibinfo{author}{Chen, X.}, \bibinfo{author}{Dai, H.H.}, \bibinfo{year}{2020}.
\newblock \bibinfo{title}{Stress-free configurations induced by a family of locally incompatible growth functions}.
\newblock \bibinfo{journal}{Journal of the Mechanics and Physics of Solids} \bibinfo{volume}{137}, \bibinfo{pages}{103834}.
\newblock \URLprefix \url{https://www.sciencedirect.com/science/article/pii/S0022509619306556}, \DOIprefix\doi{10.1016/j.jmps.2019.103834}.
\bibitem[{Cieslak et~al.(2015)Cieslak, Runions and Prusinkiewicz}]{cieslak2015auxin}
\bibinfo{author}{Cieslak, M.}, \bibinfo{author}{Runions, A.}, \bibinfo{author}{Prusinkiewicz, P.}, \bibinfo{year}{2015}.
\newblock \bibinfo{title}{Auxin-driven patterning with unidirectional fluxes}.
\newblock \bibinfo{journal}{Journal of experimental botany} \bibinfo{volume}{66}, \bibinfo{pages}{5083--5102}.
\newblock \URLprefix \url{https://academic.oup.com/jxb/article/66/16/5083/499149?login=true}, \DOIprefix\doi{10.1093/jxb/erv262}.
\bibitem[{Coen et~al.(2004)Coen, Rolland-Lagan, Matthews, Bangham and Prusinkiewicz}]{coen_genetics_2004}
\bibinfo{author}{Coen, E.}, \bibinfo{author}{Rolland-Lagan, A.G.}, \bibinfo{author}{Matthews, M.}, \bibinfo{author}{Bangham, J.A.}, \bibinfo{author}{Prusinkiewicz, P.}, \bibinfo{year}{2004}.
\newblock \bibinfo{title}{The genetics of geometry}.
\newblock \bibinfo{journal}{Proceedings of the National Academy of Sciences of the United States of America} \bibinfo{volume}{101}, \bibinfo{pages}{4728--4735}.
\newblock \URLprefix \url{http://www.pnas.org/content/101/14/4728}, \DOIprefix\doi{10.1073/pnas.0306308101}.
\bibitem[{Collinet and Lecuit(2021)}]{collinet2021}
\bibinfo{author}{Collinet, C.}, \bibinfo{author}{Lecuit, T.}, \bibinfo{year}{2021}.
\newblock \bibinfo{title}{Programmed and self-organized flow of information during morphogenesis}.
\newblock \bibinfo{journal}{Nature Reviews Molecular Cell Biology} \bibinfo{volume}{22}, \bibinfo{pages}{245--265}.
\newblock \URLprefix \url{https://www.nature.com/articles/s41580-020-00318-6}, \DOIprefix\doi{10.1038/s41580-020-00318-6}.
\bibitem[{Corson et~al.(2009)Corson, Hamant, Bohn, Traas, Boudaoud and Couder}]{corson_turning_2009}
\bibinfo{author}{Corson, F.}, \bibinfo{author}{Hamant, O.}, \bibinfo{author}{Bohn, S.}, \bibinfo{author}{Traas, J.}, \bibinfo{author}{Boudaoud, A.}, \bibinfo{author}{Couder, Y.}, \bibinfo{year}{2009}.
\newblock \bibinfo{title}{Turning a plant tissue into a living cell froth through isotropic growth}.
\newblock \bibinfo{journal}{Proceedings of the National Academy of Sciences} \bibinfo{volume}{106}, \bibinfo{pages}{8453--8458}.
\newblock \URLprefix \url{http://www.pnas.org/content/106/21/8453}, \DOIprefix\doi{10.1073/pnas.0812493106}.
\bibitem[{Cosgrove(1981)}]{cosgrove1981}
\bibinfo{author}{Cosgrove, D.J.}, \bibinfo{year}{1981}.
\newblock \bibinfo{title}{Analysis of the dynamic and steady-state responses of growth rate and turgor pressure to changes in cell parameters}.
\newblock \bibinfo{journal}{Plant Physiology} \bibinfo{volume}{68}, \bibinfo{pages}{1439--1446}.
\newblock \URLprefix \url{https://academic.oup.com/plphys/article/68/6/1439/6077837}, \DOIprefix\doi{10.1104/pp.68.6.1439}.
\bibitem[{Cosgrove(2001)}]{cosgrove_wall_2001}
\bibinfo{author}{Cosgrove, D.J.}, \bibinfo{year}{2001}.
\newblock \bibinfo{title}{Wall {Structure} and {Wall} {Loosening}. {A} {Look} {Backwards} and {Forwards}}.
\newblock \bibinfo{journal}{Plant Physiology} \bibinfo{volume}{125}, \bibinfo{pages}{131--134}.
\newblock \URLprefix \url{http://www.plantphysiol.org/content/125/1/131}, \DOIprefix\doi{10.1104/pp.125.1.131}.
\bibitem[{Cosgrove(2005)}]{cosgrove2005growth}
\bibinfo{author}{Cosgrove, D.J.}, \bibinfo{year}{2005}.
\newblock \bibinfo{title}{Growth of the plant cell wall}.
\newblock \bibinfo{journal}{Nature reviews molecular cell biology} \bibinfo{volume}{6}, \bibinfo{pages}{850--861}.
\newblock \URLprefix \url{https://www.nature.com/articles/nrm1746}, \DOIprefix\doi{10.1038/nrm1746}.
\bibitem[{Cosgrove(2018)}]{cosgrove2018diffuse}
\bibinfo{author}{Cosgrove, D.J.}, \bibinfo{year}{2018}.
\newblock \bibinfo{title}{Diffuse growth of plant cell walls}.
\newblock \bibinfo{journal}{Plant physiology} \bibinfo{volume}{176}, \bibinfo{pages}{16--27}.
\newblock \URLprefix \url{https://academic.oup.com/plphys/article-abstract/176/1/16/6116896}, \DOIprefix\doi{10.1104/pp.17.01541}.
\bibitem[{Cosgrove and Coen(2025)}]{cosgrove2025plant}
\bibinfo{author}{Cosgrove, D.J.}, \bibinfo{author}{Coen, E.}, \bibinfo{year}{2025}.
\newblock \bibinfo{title}{Plant morphogenesis: What drives growth?}
\newblock \bibinfo{journal}{Current Biology} \bibinfo{volume}{35}, \bibinfo{pages}{R283--R285}.
\newblock \URLprefix \url{https://www.cell.com/current-biology/abstract/S0960-9822(25)00263-5}, \DOIprefix\doi{j.cub.2025.02.049}.
\bibitem[{Dai and Ben~Amar(2022)}]{dai2022minimizing}
\bibinfo{author}{Dai, A.}, \bibinfo{author}{Ben~Amar, M.}, \bibinfo{year}{2022}.
\newblock \bibinfo{title}{Minimizing the elastic energy of growing leaves by conformal mapping}.
\newblock \bibinfo{journal}{Physical Review Letters} \bibinfo{volume}{129}, \bibinfo{pages}{218101}.
\newblock \URLprefix \url{https://journals.aps.org/prl/abstract/10.1103/PhysRevLett.129.218101}, \DOIprefix\doi{10.1103/PhysRevLett.129.218101}.
\bibitem[{Dainty(1963)}]{DAINTY1963279}
\bibinfo{author}{Dainty, J.}, \bibinfo{year}{1963}.
\newblock \bibinfo{title}{Water relations of plant cells}, in: \bibinfo{editor}{Preston, R.D.} (Ed.), \bibinfo{booktitle}{Volume 1}. \bibinfo{publisher}{Academic Press}. Advances in Botanical Research, pp. \bibinfo{pages}{279--326}.
\newblock \URLprefix \url{https://www.sciencedirect.com/science/article/abs/pii/S0065229608601834}, \DOIprefix\doi{10.1016/S0065-2296(08)60183-4}.
\bibitem[{Darwin(1880)}]{darwin1880power}
\bibinfo{author}{Darwin, C.}, \bibinfo{year}{1880}.
\newblock \bibinfo{title}{The power of movement in plants}.
\newblock \bibinfo{publisher}{John Murray, London}.
\bibitem[{Derr et~al.(2018)Derr, Bastien, Couturier and Douady}]{derr2018fluttering}
\bibinfo{author}{Derr, J.}, \bibinfo{author}{Bastien, R.}, \bibinfo{author}{Couturier, {\'E}.}, \bibinfo{author}{Douady, S.}, \bibinfo{year}{2018}.
\newblock \bibinfo{title}{{Fluttering of growing leaves as a way to reach flatness: experimental evidence on Persea americana}}.
\newblock \bibinfo{journal}{Journal of the Royal society interface} \bibinfo{volume}{15}, \bibinfo{pages}{20170595}.
\newblock \URLprefix \url{https://royalsocietypublishing.org/doi/full/10.1098/rsif.2017.0595}, \DOIprefix\doi{10.1098/rsif.2017.0595}.
\bibitem[{Dervaux and Ben~Amar(2008)}]{dervaux2008morphogenesis}
\bibinfo{author}{Dervaux, J.}, \bibinfo{author}{Ben~Amar, M.}, \bibinfo{year}{2008}.
\newblock \bibinfo{title}{Morphogenesis of growing soft tissues}.
\newblock \bibinfo{journal}{Physical Review Letters} \bibinfo{volume}{101}, \bibinfo{pages}{068101}.
\newblock \URLprefix \url{https://journals.aps.org/prl/abstract/10.1103/PhysRevLett.101.068101}, \DOIprefix\doi{10.1103/PhysRevLett.101.068101}.
\bibitem[{Douady and Couder(1992)}]{douady1992}
\bibinfo{author}{Douady, S.}, \bibinfo{author}{Couder, Y.}, \bibinfo{year}{1992}.
\newblock \bibinfo{title}{Phyllotaxis as a physical self-organized growth process}.
\newblock \bibinfo{journal}{Physical Review Letters} \bibinfo{volume}{68}, \bibinfo{pages}{2098}.
\newblock \URLprefix \url{https://journals.aps.org/prl/abstract/10.1103/PhysRevLett.68.2098}, \DOIprefix\doi{10.1103/PhysRevLett.68.2098}.
\bibitem[{Dumais(2021)}]{dumais2021}
\bibinfo{author}{Dumais, J.}, \bibinfo{year}{2021}.
\newblock \bibinfo{title}{Mechanics and hydraulics of pollen tube growth}.
\newblock \bibinfo{journal}{New Phytologist} \bibinfo{volume}{232}, \bibinfo{pages}{1549--1565}.
\newblock \URLprefix \url{https://nph.onlinelibrary.wiley.com/doi/10.1111/nph.17722}, \DOIprefix\doi{10.1111/nph.17722}.
\bibitem[{Dumais and Kwiatkowska(2002)}]{dumais2002}
\bibinfo{author}{Dumais, J.}, \bibinfo{author}{Kwiatkowska, D.}, \bibinfo{year}{2002}.
\newblock \bibinfo{title}{Analysis of surface growth in shoot apices}.
\newblock \bibinfo{journal}{The Plant Journal} \bibinfo{volume}{31}, \bibinfo{pages}{229--241}.
\newblock \URLprefix \url{https://onlinelibrary.wiley.com/doi/abs/10.1046/j.1365-313X.2001.01350.x}, \DOIprefix\doi{10.1046/j.1365-313X.2001.01350.x}.
\bibitem[{Dumais et~al.(2006)Dumais, Shaw, Steele, Long and Ray}]{dumais2006anisotropic}
\bibinfo{author}{Dumais, J.}, \bibinfo{author}{Shaw, S.L.}, \bibinfo{author}{Steele, C.R.}, \bibinfo{author}{Long, S.R.}, \bibinfo{author}{Ray, P.M.}, \bibinfo{year}{2006}.
\newblock \bibinfo{title}{An anisotropic-viscoplastic model of plant cell morphogenesis by tip growth}.
\newblock \bibinfo{journal}{The International journal of developmental biology} \bibinfo{volume}{50}, \bibinfo{pages}{209--222}.
\newblock \URLprefix \url{https://ijdb.ehu.eus/article/052066jd}, \DOIprefix\doi{10.1387/ijdb.052066jd}.
\bibitem[{Dupuy et~al.(2008)Dupuy, Mackenzie, Rudge and Haseloff}]{dupuy_system_2008}
\bibinfo{author}{Dupuy, L.}, \bibinfo{author}{Mackenzie, J.}, \bibinfo{author}{Rudge, T.}, \bibinfo{author}{Haseloff, J.}, \bibinfo{year}{2008}.
\newblock \bibinfo{title}{A {System} for {Modelling} {Cell}–{Cell} {Interactions} during {Plant} {Morphogenesis}}.
\newblock \bibinfo{journal}{Annals of Botany} \bibinfo{volume}{101}, \bibinfo{pages}{1255--1265}.
\newblock \URLprefix \url{https://academic.oup.com/aob/article/101/8/1255/160780}, \DOIprefix\doi{10.1093/aob/mcm235}.
\bibitem[{Dyson et~al.(2012)Dyson, Band and Jensen}]{dyson_model_2012}
\bibinfo{author}{Dyson, R.J.}, \bibinfo{author}{Band, L.R.}, \bibinfo{author}{Jensen, O.E.}, \bibinfo{year}{2012}.
\newblock \bibinfo{title}{A model of crosslink kinetics in the expanding plant cell wall: {Yield} stress and enzyme action}.
\newblock \bibinfo{journal}{Journal of Theoretical Biology} \bibinfo{volume}{307}, \bibinfo{pages}{125--136}.
\newblock \URLprefix \url{http://www.sciencedirect.com/science/article/pii/S0022519312002251}, \DOIprefix\doi{10.1016/j.jtbi.2012.04.035}.
\bibitem[{Erickson and Sax(1956)}]{erickson1956elemental}
\bibinfo{author}{Erickson, R.O.}, \bibinfo{author}{Sax, K.B.}, \bibinfo{year}{1956}.
\newblock \bibinfo{title}{Elemental growth rate of the primary root of zea mays}.
\newblock \bibinfo{journal}{Proceedings of the American Philosophical Society} \bibinfo{volume}{100}, \bibinfo{pages}{487--498}.
\newblock \URLprefix \url{https://www.jstor.org/stable/3143682}.
\bibitem[{Erickson and Silk(1980)}]{erickson1980}
\bibinfo{author}{Erickson, R.O.}, \bibinfo{author}{Silk, W.K.}, \bibinfo{year}{1980}.
\newblock \bibinfo{title}{The kinematics of plant growth}.
\newblock \bibinfo{journal}{Scientific American} \bibinfo{volume}{242}, \bibinfo{pages}{134--151}.
\newblock \URLprefix \url{http://www.jstor.org/stable/24966328}.
\bibitem[{Fishman and Génard(1998)}]{fishman_genard_1998}
\bibinfo{author}{Fishman, S.}, \bibinfo{author}{Génard, M.}, \bibinfo{year}{1998}.
\newblock \bibinfo{title}{A biophysical model of fruit growth: simulation of seasonal and diurnal dynamics of mass}.
\newblock \bibinfo{journal}{Plant, Cell \& Environment} \bibinfo{volume}{21}, \bibinfo{pages}{739--752}.
\newblock \URLprefix \url{https://onlinelibrary.wiley.com/doi/full/10.1046/j.1365-3040.1998.00322.x}, \DOIprefix\doi{10.1046/j.1365-3040.1998.00322.x}.
\bibitem[{Forterre(2022)}]{forterre2022basic}
\bibinfo{author}{Forterre, Y.}, \bibinfo{year}{2022}.
\newblock \bibinfo{title}{Basic soft matter for plants}, in: \bibinfo{editor}{Jensen, K.}, \bibinfo{editor}{Forterre, Y.} (Eds.), \bibinfo{booktitle}{Soft Matter in Plants: From Biophysics to Biomimetics}. \bibinfo{publisher}{The Royal Society of Chemistry}. number~\bibinfo{number}{15} in \bibinfo{series}{Soft Matter Series}. chapter~\bibinfo{chapter}{1}, pp. \bibinfo{pages}{1 -- 65}.
\newblock \URLprefix \url{https://books.rsc.org/books/edited-volume/2002/chapter/4582122}, \DOIprefix\doi{10.1039/9781839161162-00001}.
\bibitem[{Fozard et~al.(2013)Fozard, Lucas, King and Jensen}]{fozard_vertex-element_2013}
\bibinfo{author}{Fozard, J.A.}, \bibinfo{author}{Lucas, M.}, \bibinfo{author}{King, J.R.}, \bibinfo{author}{Jensen, O.E.}, \bibinfo{year}{2013}.
\newblock \bibinfo{title}{Vertex-element models for anisotropic growth of elongated plant organs}.
\newblock \bibinfo{journal}{Front Plant Sci} \bibinfo{volume}{4}.
\newblock \URLprefix \url{http://www.ncbi.nlm.nih.gov/pmc/articles/PMC3706750/}, \DOIprefix\doi{10.3389/fpls.2013.00233}.
\bibitem[{Fruleux and Boudaoud(2019)}]{fruleux2019modulation}
\bibinfo{author}{Fruleux, A.}, \bibinfo{author}{Boudaoud, A.}, \bibinfo{year}{2019}.
\newblock \bibinfo{title}{Modulation of tissue growth heterogeneity by responses to mechanical stress}.
\newblock \bibinfo{journal}{Proceedings of the National Academy of Sciences} \bibinfo{volume}{116}, \bibinfo{pages}{1940--1945}.
\newblock \URLprefix \url{https://www.pnas.org/doi/10.1073/pnas.1815342116}, \DOIprefix\doi{10.1073/pnas.1815342116}.
\bibitem[{Gayon(2000)}]{gayon2000history}
\bibinfo{author}{Gayon, J.}, \bibinfo{year}{2000}.
\newblock \bibinfo{title}{History of the concept of allometry}.
\newblock \bibinfo{journal}{American zoologist} \bibinfo{volume}{40}, \bibinfo{pages}{748--758}.
\newblock \URLprefix \url{https://academic.oup.com/icb/article/40/5/748/157095}, \DOIprefix\doi{10.1093/icb/40.5.748}.
\bibitem[{Ghysels et~al.(2010)Ghysels, Samaey, Van~Liedekerke, Tijskens, Ramon and Roose}]{ghysels2010multiscale}
\bibinfo{author}{Ghysels, P.}, \bibinfo{author}{Samaey, G.}, \bibinfo{author}{Van~Liedekerke, P.}, \bibinfo{author}{Tijskens, E.}, \bibinfo{author}{Ramon, H.}, \bibinfo{author}{Roose, D.}, \bibinfo{year}{2010}.
\newblock \bibinfo{title}{Multiscale modeling of viscoelastic plant tissue}.
\newblock \bibinfo{journal}{International Journal for Multiscale Computational Engineering} \bibinfo{volume}{8}.
\newblock \URLprefix \url{https://www.dl.begellhouse.com/journals/61fd1b191cf7e96f}, \DOIprefix\doi{10.1615/IntJMultCompEng.v8.i4.30}.
\bibitem[{Godin et~al.(2020)Godin, Gol{\'e} and Douady}]{godin2020phyllotaxis}
\bibinfo{author}{Godin, C.}, \bibinfo{author}{Gol{\'e}, C.}, \bibinfo{author}{Douady, S.}, \bibinfo{year}{2020}.
\newblock \bibinfo{title}{Phyllotaxis as geometric canalization during plant development}.
\newblock \bibinfo{journal}{Development} \bibinfo{volume}{147}, \bibinfo{pages}{dev165878}.
\newblock \URLprefix \url{https://journals.biologists.com/dev/article/147/19/dev165878/225951/}, \DOIprefix\doi{10.1242/dev.165878}.
\bibitem[{Goriely(2017)}]{Goriely2017}
\bibinfo{author}{Goriely, A.}, \bibinfo{year}{2017}.
\newblock \bibinfo{title}{{The Mathematics and Mechanics of Biological Growth}}. volume~\bibinfo{volume}{45} of \textit{\bibinfo{series}{Interdisciplinary applied mathematics}}.
\newblock \bibinfo{publisher}{Springer-Verlag, New York}.
\newblock \URLprefix \url{https://link.springer.com/book/10.1007/978-0-387-87710-5}, \DOIprefix\doi{10.1007/978-0-387-87710-5}.
\bibitem[{Goriely and Ben~Amar(2005)}]{goriely2005morpho}
\bibinfo{author}{Goriely, A.}, \bibinfo{author}{Ben~Amar, M.}, \bibinfo{year}{2005}.
\newblock \bibinfo{title}{Morpho-{\'e}lasticit{\'e}}, in: \bibinfo{editor}{Chazottes, J.R.}, \bibinfo{editor}{Joets, A.}, \bibinfo{editor}{Letellier, C.}, \bibinfo{editor}{Ribotta, R.} (Eds.), \bibinfo{booktitle}{Compte-rendus de la 8\textsuperscript{e} Rencontre du Non-Lin\'eaire}, \bibinfo{address}{Paris}. p.~\bibinfo{pages}{97}.
\newblock \URLprefix \url{http://nonlineaire.univ-lille1.fr/SNL/media/pnl_archive/ftp/2005/C-Rendus/Actes05.pdf}.
\bibitem[{Goriely and Tabor(1998)}]{goriely1998spontaneous}
\bibinfo{author}{Goriely, A.}, \bibinfo{author}{Tabor, M.}, \bibinfo{year}{1998}.
\newblock \bibinfo{title}{Spontaneous helix hand reversal and tendril perversion in climbing plants}.
\newblock \bibinfo{journal}{Physical Review Letters} \bibinfo{volume}{80}, \bibinfo{pages}{1564}.
\newblock \URLprefix \url{https://journals.aps.org/prl/abstract/10.1103/PhysRevLett.80.1564}, \DOIprefix\doi{10.1103/PhysRevLett.80.1564}.
\bibitem[{Grieneisen et~al.(2007)Grieneisen, Xu, Mar{\'e}e, Hogeweg and Scheres}]{grieneisen2007auxin}
\bibinfo{author}{Grieneisen, V.A.}, \bibinfo{author}{Xu, J.}, \bibinfo{author}{Mar{\'e}e, A.F.}, \bibinfo{author}{Hogeweg, P.}, \bibinfo{author}{Scheres, B.}, \bibinfo{year}{2007}.
\newblock \bibinfo{title}{Auxin transport is sufficient to generate a maximum and gradient guiding root growth}.
\newblock \bibinfo{journal}{Nature} \bibinfo{volume}{449}, \bibinfo{pages}{1008}.
\newblock \URLprefix \url{https://www.nature.com/articles/nature06215}, \DOIprefix\doi{10.1038/nature06215}.
\bibitem[{Guo et~al.(2025)Guo, Zhang, Paradiso, Long, Hsia and Liu}]{guo2025midveins}
\bibinfo{author}{Guo, K.}, \bibinfo{author}{Zhang, Y.}, \bibinfo{author}{Paradiso, M.}, \bibinfo{author}{Long, Y.}, \bibinfo{author}{Hsia, K.J.}, \bibinfo{author}{Liu, M.}, \bibinfo{year}{2025}.
\newblock \bibinfo{title}{Midveins regulate the shape formation of drying leaves}.
\newblock \URLprefix \url{https://arxiv.org/abs/2507.01813}, \DOIprefix\doi{10.48550/arXiv.2507.01813}.
\bibitem[{Halder et~al.(1995)Halder, Callaerts and Gehring}]{halder1995}
\bibinfo{author}{Halder, G.}, \bibinfo{author}{Callaerts, P.}, \bibinfo{author}{Gehring, W.J.}, \bibinfo{year}{1995}.
\newblock \bibinfo{title}{Induction of ectopic eyes by targeted expression of the eyeless gene in \textit{Drosophila}}.
\newblock \bibinfo{journal}{Science} \bibinfo{volume}{267}, \bibinfo{pages}{1788--1792}.
\newblock \URLprefix \url{https://www.science.org/doi/10.1126/science.7892602}, \DOIprefix\doi{10.1126/science.7892602}.
\bibitem[{Hamant et~al.(2008)Hamant, Heisler, J\"{o}nsson, Krupinski, Uyttewaal, Bokov, Corson, Sahlin, Boudaoud, Meyerowitz, Couder and Traas}]{hamant_developmental_2008}
\bibinfo{author}{Hamant, O.}, \bibinfo{author}{Heisler, M.G.}, \bibinfo{author}{J\"{o}nsson, H.}, \bibinfo{author}{Krupinski, P.}, \bibinfo{author}{Uyttewaal, M.}, \bibinfo{author}{Bokov, P.}, \bibinfo{author}{Corson, F.}, \bibinfo{author}{Sahlin, P.}, \bibinfo{author}{Boudaoud, A.}, \bibinfo{author}{Meyerowitz, E.M.}, \bibinfo{author}{Couder, Y.}, \bibinfo{author}{Traas, J.}, \bibinfo{year}{2008}.
\newblock \bibinfo{title}{Developmental {Patterning} by {Mechanical} {Signals} in {Arabidopsis}}.
\newblock \bibinfo{journal}{Science} \bibinfo{volume}{322}, \bibinfo{pages}{1650--1655}.
\newblock \URLprefix \url{http://www.sciencemag.org/content/322/5908/1650}, \DOIprefix\doi{10.1126/science.1165594}.
\bibitem[{Hamant and Traas(2010)}]{hamant_mechanics_2010}
\bibinfo{author}{Hamant, O.}, \bibinfo{author}{Traas, J.}, \bibinfo{year}{2010}.
\newblock \bibinfo{title}{The mechanics behind plant development}.
\newblock \bibinfo{journal}{New Phytologist} \bibinfo{volume}{185}, \bibinfo{pages}{369--385}.
\newblock \URLprefix \url{http://onlinelibrary.wiley.com/doi/10.1111/j.1469-8137.2009.03100.x/abstract}, \DOIprefix\doi{10.1111/j.1469-8137.2009.03100.x}.
\bibitem[{Hartmann et~al.(2019)Hartmann, Barbier~de Reuille and Kuhlemeier}]{hartmann2019toward}
\bibinfo{author}{Hartmann, F.P.}, \bibinfo{author}{Barbier~de Reuille, P.}, \bibinfo{author}{Kuhlemeier, C.}, \bibinfo{year}{2019}.
\newblock \bibinfo{title}{{Toward a 3D model of phyllotaxis based on a biochemically plausible auxin-transport mechanism}}.
\newblock \bibinfo{journal}{PLoS computational biology} \bibinfo{volume}{15}, \bibinfo{pages}{e1006896}.
\newblock \URLprefix \url{https://journals.plos.org/ploscompbiol/article?id=10.1371/journal.pcbi.1006896}, \DOIprefix\doi{10.1371/journal.pcbi.1006896}.
\bibitem[{Heisler et~al.(2010)Heisler, Hamant, Krupinski, Uyttewaal, Ohno, J{\"o}nsson, Traas and Meyerowitz}]{heisler2010alignment}
\bibinfo{author}{Heisler, M.G.}, \bibinfo{author}{Hamant, O.}, \bibinfo{author}{Krupinski, P.}, \bibinfo{author}{Uyttewaal, M.}, \bibinfo{author}{Ohno, C.}, \bibinfo{author}{J{\"o}nsson, H.}, \bibinfo{author}{Traas, J.}, \bibinfo{author}{Meyerowitz, E.M.}, \bibinfo{year}{2010}.
\newblock \bibinfo{title}{{Alignment between PIN1 polarity and microtubule orientation in the shoot apical meristem reveals a tight coupling between morphogenesis and auxin transport}}.
\newblock \bibinfo{journal}{PLoS biology} \bibinfo{volume}{8}, \bibinfo{pages}{e1000516}.
\newblock \URLprefix \url{https://journals.plos.org/plosbiology/article?id=10.1371/journal.pbio.1000516}, \DOIprefix\doi{10.1371/journal.pbio.1000516}.
\bibitem[{Hejnowicz and Romberger(1984)}]{hejnowicz1984growth}
\bibinfo{author}{Hejnowicz, Z.}, \bibinfo{author}{Romberger, J.A.}, \bibinfo{year}{1984}.
\newblock \bibinfo{title}{Growth tensor of plant organs}.
\newblock \bibinfo{journal}{Journal of Theoretical Biology} \bibinfo{volume}{110}, \bibinfo{pages}{93--114}.
\newblock \URLprefix \url{https://www.sciencedirect.com/science/article/pii/S002251938480017X}, \DOIprefix\doi{10.1016/S0022-5193(84)80017-X}.
\bibitem[{Hern{\'a}ndez-Hern{\'a}ndez et~al.(2024)Hern{\'a}ndez-Hern{\'a}ndez, Marchand, Kiss and Boudaoud}]{hernandez2024}
\bibinfo{author}{Hern{\'a}ndez-Hern{\'a}ndez, V.}, \bibinfo{author}{Marchand, O.C.}, \bibinfo{author}{Kiss, A.}, \bibinfo{author}{Boudaoud, A.}, \bibinfo{year}{2024}.
\newblock \bibinfo{title}{A mechanohydraulic model supports a role for plasmodesmata in cotton fiber elongation}.
\newblock \bibinfo{journal}{PNAS nexus} \bibinfo{volume}{3}, \bibinfo{pages}{pgae256}.
\newblock \URLprefix \url{https://academic.oup.com/pnasnexus/article/3/7/pgae256/7713014}, \DOIprefix\doi{10.1093/pnasnexus/pgae256}.
\bibitem[{Hervieux et~al.(2016)Hervieux, Dumond, Sapala, Routier-Kierzkowska, Kierzkowski, Roeder, Smith, Boudaoud and Hamant}]{hervieux2016mechanical}
\bibinfo{author}{Hervieux, N.}, \bibinfo{author}{Dumond, M.}, \bibinfo{author}{Sapala, A.}, \bibinfo{author}{Routier-Kierzkowska, A.L.}, \bibinfo{author}{Kierzkowski, D.}, \bibinfo{author}{Roeder, A.H.}, \bibinfo{author}{Smith, R.S.}, \bibinfo{author}{Boudaoud, A.}, \bibinfo{author}{Hamant, O.}, \bibinfo{year}{2016}.
\newblock \bibinfo{title}{A mechanical feedback restricts sepal growth and shape in arabidopsis}.
\newblock \bibinfo{journal}{Current Biology} \bibinfo{volume}{26}, \bibinfo{pages}{1019--1028}.
\newblock \URLprefix \url{https://www.sciencedirect.com/science/article/pii/S0960982216301804}, \DOIprefix\doi{10.1016/j.cub.2016.03.004}.
\bibitem[{Holzapfel(2000)}]{holzapfel2000nonlinear}
\bibinfo{author}{Holzapfel, G.A.}, \bibinfo{year}{2000}.
\newblock \bibinfo{title}{Nonlinear solid mechanics}.
\newblock \bibinfo{publisher}{John Wiley \& Sons Ltd.}, \bibinfo{address}{Chichester}.
\newblock \URLprefix \url{https://www.wiley.com/en-us/Nonlinear+Solid+Mechanics\%3A+A+Continuum+Approach+for+Engineering-p-9780471823193}.
\bibitem[{Huang et~al.(2018)Huang, Wang, Quinn, Suresh and Hsia}]{huang2018differential}
\bibinfo{author}{Huang, C.}, \bibinfo{author}{Wang, Z.}, \bibinfo{author}{Quinn, D.}, \bibinfo{author}{Suresh, S.}, \bibinfo{author}{Hsia, K.J.}, \bibinfo{year}{2018}.
\newblock \bibinfo{title}{Differential growth and shape formation in plant organs}.
\newblock \bibinfo{journal}{Proceedings of the National Academy of Sciences} \bibinfo{volume}{115}, \bibinfo{pages}{12359--12364}.
\newblock \URLprefix \url{http://www.pnas.org/cgi/doi/10.1073/pnas.1811296115}, \DOIprefix\doi{10.1073/pnas.1811296115}.
\bibitem[{Huxley(1924)}]{huxley1924constant}
\bibinfo{author}{Huxley, J.S.}, \bibinfo{year}{1924}.
\newblock \bibinfo{title}{Constant differential growth-ratios and their significance}.
\newblock \bibinfo{journal}{Nature} \bibinfo{volume}{114}, \bibinfo{pages}{895--896}.
\newblock \URLprefix \url{https://www.nature.com/articles/114895a0}, \DOIprefix\doi{10.1038/114895a0}.
\bibitem[{Huxley(1932)}]{huxley1993problems}
\bibinfo{author}{Huxley, J.S.}, \bibinfo{year}{1932}.
\newblock \bibinfo{title}{Problems of relative growth}.
\newblock \bibinfo{publisher}{L. MacVeagh, The Dial Press}, \bibinfo{address}{New York}.
\newblock \URLprefix \url{https://www.biodiversitylibrary.org/bibliography/6427}, \DOIprefix\doi{10.5962/bhl.title.6427}.
\bibitem[{Huxley and Teissier(1936)}]{huxley1936terminology}
\bibinfo{author}{Huxley, J.S.}, \bibinfo{author}{Teissier, G.}, \bibinfo{year}{1936}.
\newblock \bibinfo{title}{Terminology of relative growth}.
\newblock \bibinfo{journal}{Nature} \bibinfo{volume}{137}, \bibinfo{pages}{780--781}.
\newblock \URLprefix \url{https://www.nature.com/articles/137780b0}, \DOIprefix\doi{10.1038/137780b0}.
\bibitem[{Jones and Chapman(2012)}]{jones2012modeling}
\bibinfo{author}{Jones, G.W.}, \bibinfo{author}{Chapman, S.J.}, \bibinfo{year}{2012}.
\newblock \bibinfo{title}{Modeling growth in biological materials}.
\newblock \bibinfo{journal}{{SIAM Review}} \bibinfo{volume}{54}, \bibinfo{pages}{52--118}.
\newblock \URLprefix \url{https://epubs.siam.org/doi/abs/10.1137/080731785}, \DOIprefix\doi{10.1137/080731785}.
\bibitem[{J\"onsson et~al.(2006)J\"onsson, Heisler, Shapiro, Meyerowitz and Mjolsness}]{jonsson_auxin-driven_2006}
\bibinfo{author}{J\"onsson, H.}, \bibinfo{author}{Heisler, M.G.}, \bibinfo{author}{Shapiro, B.E.}, \bibinfo{author}{Meyerowitz, E.M.}, \bibinfo{author}{Mjolsness, E.}, \bibinfo{year}{2006}.
\newblock \bibinfo{title}{An auxin-driven polarized transport model for phyllotaxis}.
\newblock \bibinfo{journal}{Proceedings of the National Academy of Sciences} \bibinfo{volume}{103}, \bibinfo{pages}{1633--1638}.
\newblock \URLprefix \url{http://www.pnas.org/content/103/5/1633}, \DOIprefix\doi{10.1073/pnas.0509839103}.
\bibitem[{J{\"u}licher et~al.(2018)J{\"u}licher, Grill and Salbreux}]{julicher2018hydrodynamic}
\bibinfo{author}{J{\"u}licher, F.}, \bibinfo{author}{Grill, S.W.}, \bibinfo{author}{Salbreux, G.}, \bibinfo{year}{2018}.
\newblock \bibinfo{title}{Hydrodynamic theory of active matter}.
\newblock \bibinfo{journal}{Reports on Progress in Physics} \bibinfo{volume}{81}, \bibinfo{pages}{076601}.
\newblock \URLprefix \url{https://iopscience.iop.org/article/10.1088/1361-6633/aab6bb/meta}, \DOIprefix\doi{10.1088/1361-6633/aab6bb}.
\bibitem[{Kaplan and Hagemann(1991)}]{kaplan1991relationship}
\bibinfo{author}{Kaplan, D.R.}, \bibinfo{author}{Hagemann, W.}, \bibinfo{year}{1991}.
\newblock \bibinfo{title}{The relationship of cell and organism in vascular plants}.
\newblock \bibinfo{journal}{Bioscience} \bibinfo{volume}{41}, \bibinfo{pages}{693--703}.
\newblock \URLprefix \url{https://www.jstor.org/stable/1311764}, \DOIprefix\doi{10.2307/1311764}.
\bibitem[{Kavanagh and Richards(1942)}]{kavanagh1942mathematical}
\bibinfo{author}{Kavanagh, A.J.}, \bibinfo{author}{Richards, O.W.}, \bibinfo{year}{1942}.
\newblock \bibinfo{title}{Mathematical analysis of the relative growth of organisms}.
\newblock \bibinfo{journal}{Proceedings of the Rochester Academy of Science} \bibinfo{volume}{8}, \bibinfo{pages}{150--174}.
\newblock \URLprefix \url{https://nyheritage.contentdm.oclc.org/digital/collection/p16694coll84/id/6037/rec/1}.
\bibitem[{Kennaway et~al.(2011)Kennaway, Coen, Green and Bangham}]{kennaway_generation_2011}
\bibinfo{author}{Kennaway, R.}, \bibinfo{author}{Coen, E.}, \bibinfo{author}{Green, A.}, \bibinfo{author}{Bangham, A.}, \bibinfo{year}{2011}.
\newblock \bibinfo{title}{Generation of {Diverse} {Biological} {Forms} through {Combinatorial} {Interactions} between {Tissue} {Polarity} and {Growth}}.
\newblock \bibinfo{journal}{PLOS Comput Biol} \bibinfo{volume}{7}, \bibinfo{pages}{e1002071}.
\newblock \URLprefix \url{http://journals.plos.org/ploscompbiol/article?id=10.1371/journal.pcbi.1002071}, \DOIprefix\doi{10.1371/journal.pcbi.1002071}.
\bibitem[{Khadka et~al.(2019)Khadka, Julien and Alim}]{khadka2019feedback}
\bibinfo{author}{Khadka, J.}, \bibinfo{author}{Julien, J.D.}, \bibinfo{author}{Alim, K.}, \bibinfo{year}{2019}.
\newblock \bibinfo{title}{Feedback from tissue mechanics self-organizes efficient outgrowth of plant organ}.
\newblock \bibinfo{journal}{Biophysical journal} \bibinfo{volume}{117}, \bibinfo{pages}{1995--2004}.
\newblock \URLprefix \url{https://www.sciencedirect.com/science/article/pii/S0006349519308677}, \DOIprefix\doi{10.1016/j.bpj.2019.10.019}.
\bibitem[{Kierzkowski et~al.(2012)Kierzkowski, Nakayama, Routier-Kierzkowska, Weber, Bayer, Schorderet, Reinhardt, Kuhlemeier and Smith}]{kierzkowski_elastic_2012}
\bibinfo{author}{Kierzkowski, D.}, \bibinfo{author}{Nakayama, N.}, \bibinfo{author}{Routier-Kierzkowska, A.L.}, \bibinfo{author}{Weber, A.}, \bibinfo{author}{Bayer, E.}, \bibinfo{author}{Schorderet, M.}, \bibinfo{author}{Reinhardt, D.}, \bibinfo{author}{Kuhlemeier, C.}, \bibinfo{author}{Smith, R.S.}, \bibinfo{year}{2012}.
\newblock \bibinfo{title}{Elastic domains regulate growth and organogenesis in the plant shoot apical meristem}.
\newblock \bibinfo{journal}{Science} \bibinfo{volume}{335}, \bibinfo{pages}{1096--1099}.
\newblock \URLprefix \url{http://science.sciencemag.org/content/335/6072/1096}, \DOIprefix\doi{10.1126/science.1213100}.
\bibitem[{Kramer and Boyer(1995)}]{kramer1995water}
\bibinfo{author}{Kramer, P.J.}, \bibinfo{author}{Boyer, J.S.}, \bibinfo{year}{1995}.
\newblock \bibinfo{title}{Water relations of plants and soils}.
\newblock \bibinfo{publisher}{Academic press}, \bibinfo{address}{New York}.
\bibitem[{Kutschera(1989)}]{kutschera1989tissue}
\bibinfo{author}{Kutschera, U.}, \bibinfo{year}{1989}.
\newblock \bibinfo{title}{Tissue stresses in growing plant organs}.
\newblock \bibinfo{journal}{Physiologia Plantarum} \bibinfo{volume}{77}, \bibinfo{pages}{157--163}.
\newblock \URLprefix \url{https://link.springer.com/article/10.1007/BF00397885}, \DOIprefix\doi{10.1111/j.1399-3054.1989.tb05992.x}.
\bibitem[{Kutschera and Niklas(2007)}]{kutschera_epidermal-growth-control_2007}
\bibinfo{author}{Kutschera, U.}, \bibinfo{author}{Niklas, K.J.}, \bibinfo{year}{2007}.
\newblock \bibinfo{title}{The epidermal-growth-control theory of stem elongation: An old and a new perspective}.
\newblock \bibinfo{journal}{Journal of Plant Physiology} \bibinfo{volume}{164}, \bibinfo{pages}{1395--1409}.
\newblock \URLprefix \url{http://www.sciencedirect.com/science/article/pii/S017616170700226X}, \DOIprefix\doi{10.1016/j.jplph.2007.08.002}.
\bibitem[{Kwiatkowska and Dumais(2003)}]{kwiatkowska2003growth}
\bibinfo{author}{Kwiatkowska, D.}, \bibinfo{author}{Dumais, J.}, \bibinfo{year}{2003}.
\newblock \bibinfo{title}{{Growth and morphogenesis at the vegetative shoot apex of Anagallis arvensis L.}}
\newblock \bibinfo{journal}{Journal of Experimental Botany} \bibinfo{volume}{54}, \bibinfo{pages}{1585--1595}.
\newblock \URLprefix \url{https://academic.oup.com/jxb/article/54/387/1585/540348}, \DOIprefix\doi{10.1093/jxb/erg166}.
\bibitem[{Laplaud et~al.(2024)Laplaud, Muller, Demidova, Drevensek and Boudaoud}]{laplaud2024assessing}
\bibinfo{author}{Laplaud, V.}, \bibinfo{author}{Muller, E.}, \bibinfo{author}{Demidova, N.}, \bibinfo{author}{Drevensek, S.}, \bibinfo{author}{Boudaoud, A.}, \bibinfo{year}{2024}.
\newblock \bibinfo{title}{Assessing the hydromechanical control of plant growth}.
\newblock \bibinfo{journal}{Journal of the Royal Society Interface} \bibinfo{volume}{21}, \bibinfo{pages}{20240008}.
\newblock \URLprefix \url{https://royalsocietypublishing.org/doi/abs/10.1098/rsif.2024.0008}, \DOIprefix\doi{10.1098/rsif.2024.0008}.
\bibitem[{Lapointe et~al.(2025)Lapointe, Kaur, Routier-Kierzkowska and Burian}]{LAPOINTE2025102759}
\bibinfo{author}{Lapointe, B.P.}, \bibinfo{author}{Kaur, N.S.}, \bibinfo{author}{Routier-Kierzkowska, A.L.}, \bibinfo{author}{Burian, A.}, \bibinfo{year}{2025}.
\newblock \bibinfo{title}{From stress to growth: Mechanical tissue interactions in developing organs}.
\newblock \bibinfo{journal}{Current Opinion in Plant Biology} \bibinfo{volume}{86}, \bibinfo{pages}{102759}.
\newblock \URLprefix \url{https://www.sciencedirect.com/science/article/pii/S1369526625000731}, \DOIprefix\doi{10.1016/j.pbi.2025.102759}.
\bibitem[{Levine and Goldman(2023)}]{levine2023physics}
\bibinfo{author}{Levine, H.}, \bibinfo{author}{Goldman, D.I.}, \bibinfo{year}{2023}.
\newblock \bibinfo{title}{Physics of smart active matter: integrating active matter and control to gain insights into living systems}.
\newblock \bibinfo{journal}{Soft Matter} \bibinfo{volume}{19}, \bibinfo{pages}{4204--4207}.
\newblock \URLprefix \url{https://pubs.rsc.org/en/content/articlelanding/2023/sm/d3sm00171g}, \DOIprefix\doi{10.1039/d3sm00171g}.
\bibitem[{Lewicka and Pietruszka(2007)}]{lewicka2007anisotropic}
\bibinfo{author}{Lewicka, S.}, \bibinfo{author}{Pietruszka, M.}, \bibinfo{year}{2007}.
\newblock \bibinfo{title}{Anisotropic plant cell elongation due to ortho-gravitropism}.
\newblock \bibinfo{journal}{Journal of mathematical biology} \bibinfo{volume}{54}, \bibinfo{pages}{91--100}.
\newblock \URLprefix \url{https://link.springer.com/article/10.1007/s00285-006-0049-3}, \DOIprefix\doi{10.1007/s00285-006-0049-3}.
\bibitem[{Liang and Mahadevan(2009)}]{liang2009shape}
\bibinfo{author}{Liang, H.}, \bibinfo{author}{Mahadevan, L.}, \bibinfo{year}{2009}.
\newblock \bibinfo{title}{The shape of a long leaf}.
\newblock \bibinfo{journal}{Proceedings of the National Academy of Sciences} \bibinfo{volume}{106}, \bibinfo{pages}{22049--22054}.
\newblock \URLprefix \url{https://www.pnas.org/doi/10.1073/pnas.0911954106}, \DOIprefix\doi{10.1073/pnas.0911954106}.
\bibitem[{Lockhart(1965)}]{lockhart1965analysis}
\bibinfo{author}{Lockhart, J.A.}, \bibinfo{year}{1965}.
\newblock \bibinfo{title}{An analysis of irreversible plant cell elongation}.
\newblock \bibinfo{journal}{Journal of theoretical biology} \bibinfo{volume}{8}, \bibinfo{pages}{264--275}.
\newblock \URLprefix \url{https://www.sciencedirect.com/science/article/pii/0022519365900779}, \DOIprefix\doi{10.1016/0022-5193(65)90077-9}.
\bibitem[{Long et~al.(2020)Long, Cheddadi, Mosca, Mirabet, Dumond, Kiss, Traas, Godin and Boudaoud}]{long2020cellular}
\bibinfo{author}{Long, Y.}, \bibinfo{author}{Cheddadi, I.}, \bibinfo{author}{Mosca, G.}, \bibinfo{author}{Mirabet, V.}, \bibinfo{author}{Dumond, M.}, \bibinfo{author}{Kiss, A.}, \bibinfo{author}{Traas, J.}, \bibinfo{author}{Godin, C.}, \bibinfo{author}{Boudaoud, A.}, \bibinfo{year}{2020}.
\newblock \bibinfo{title}{Cellular heterogeneity in pressure and growth emerges from tissue topology and geometry}.
\newblock \bibinfo{journal}{Current Biology} \bibinfo{volume}{30}, \bibinfo{pages}{1504--1516}.
\newblock \URLprefix \url{https://www.sciencedirect.com/science/article/pii/S0960982220302001}, \DOIprefix\doi{10.1016/j.cub.2020.02.027}.
\bibitem[{Marchetti et~al.(2013)Marchetti, Joanny, Ramaswamy, Liverpool, Prost, Rao and Simha}]{marchetti2013hydrodynamics}
\bibinfo{author}{Marchetti, M.C.}, \bibinfo{author}{Joanny, J.F.}, \bibinfo{author}{Ramaswamy, S.}, \bibinfo{author}{Liverpool, T.B.}, \bibinfo{author}{Prost, J.}, \bibinfo{author}{Rao, M.}, \bibinfo{author}{Simha, R.A.}, \bibinfo{year}{2013}.
\newblock \bibinfo{title}{Hydrodynamics of soft active matter}.
\newblock \bibinfo{journal}{Reviews of modern physics} \bibinfo{volume}{85}, \bibinfo{pages}{1143--1189}.
\newblock \URLprefix \url{https://journals.aps.org/rmp/abstract/10.1103/RevModPhys.85.1143}, \DOIprefix\doi{10.1103/RevModPhys.85.1143}.
\bibitem[{Meinhardt et~al.(1998)Meinhardt, Koch and Bernasconi}]{meinhardt1998models}
\bibinfo{author}{Meinhardt, H.}, \bibinfo{author}{Koch, A.J.}, \bibinfo{author}{Bernasconi, G.}, \bibinfo{year}{1998}.
\newblock \bibinfo{title}{Models of pattern formation applied to plant development}, in: \bibinfo{editor}{Barabe, D.}, \bibinfo{editor}{Jean, R.V.} (Eds.), \bibinfo{booktitle}{Symmetry in plants}. \bibinfo{publisher}{Singapore: World Scientific}. volume~\bibinfo{volume}{4}, pp. \bibinfo{pages}{723--758}.
\bibitem[{Merks et~al.(2011)Merks, Guravage, Inz{\'e} and Beemster}]{merks2011virtualleaf}
\bibinfo{author}{Merks, R.M.}, \bibinfo{author}{Guravage, M.}, \bibinfo{author}{Inz{\'e}, D.}, \bibinfo{author}{Beemster, G.T.}, \bibinfo{year}{2011}.
\newblock \bibinfo{title}{Virtual{L}eaf: an open-source framework for cell-based modeling of plant tissue growth and development}.
\newblock \bibinfo{journal}{Plant physiology} \bibinfo{volume}{155}, \bibinfo{pages}{656--666}.
\newblock \URLprefix \url{https://academic.oup.com/plphys/article/155/2/656/6111407}, \DOIprefix\doi{10.1104/pp.110.167619}.
\bibitem[{Mietke et~al.(2019)Mietke, Jemseena, Kumar, Sbalzarini and J\"ulicher}]{PhysRevLett.123.188101}
\bibinfo{author}{Mietke, A.}, \bibinfo{author}{Jemseena, V.}, \bibinfo{author}{Kumar, K.V.}, \bibinfo{author}{Sbalzarini, I.F.}, \bibinfo{author}{J\"ulicher, F.}, \bibinfo{year}{2019}.
\newblock \bibinfo{title}{Minimal model of cellular symmetry breaking}.
\newblock \bibinfo{journal}{Phys. Rev. Lett.} \bibinfo{volume}{123}, \bibinfo{pages}{188101}.
\newblock \URLprefix \url{https://link.aps.org/doi/10.1103/PhysRevLett.123.188101}, \DOIprefix\doi{10.1103/PhysRevLett.123.188101}.
\bibitem[{Molz and Boyer(1978)}]{molz1978growth}
\bibinfo{author}{Molz, F.J.}, \bibinfo{author}{Boyer, J.S.}, \bibinfo{year}{1978}.
\newblock \bibinfo{title}{Growth-induced water potentials in plant cells and tissues}.
\newblock \bibinfo{journal}{Plant Physiology} \bibinfo{volume}{62}, \bibinfo{pages}{423--429}.
\newblock \URLprefix \url{https://academic.oup.com/plphys/article/62/3/423/6075923}, \DOIprefix\doi{10.1104/pp.62.3.423}.
\bibitem[{Molz and Ikenberry(1974)}]{molz1974water}
\bibinfo{author}{Molz, F.J.}, \bibinfo{author}{Ikenberry, E.}, \bibinfo{year}{1974}.
\newblock \bibinfo{title}{Water transport through plant cells and cell walls: theoretical development}.
\newblock \bibinfo{journal}{Soil Science Society of America Journal} \bibinfo{volume}{38}, \bibinfo{pages}{699--704}.
\newblock \URLprefix \url{https://acsess.onlinelibrary.wiley.com/doi/abs/10.2136/sssaj1974.03615995003800050009x}, \DOIprefix\doi{10.2136/sssaj1974.03615995003800050009x}.
\bibitem[{Molz et~al.(1975)Molz, Truelove and Peterson}]{molz1975dynamics}
\bibinfo{author}{Molz, F.J.}, \bibinfo{author}{Truelove, B.}, \bibinfo{author}{Peterson, C.M.}, \bibinfo{year}{1975}.
\newblock \bibinfo{title}{{Dynamics of Rehydration in Leaf Disks}}.
\newblock \bibinfo{journal}{Agronomy Journal} \bibinfo{volume}{67}, \bibinfo{pages}{511--515}.
\newblock \URLprefix \url{https://acsess.onlinelibrary.wiley.com/doi/10.2134/agronj1975.00021962006700040014x}, \DOIprefix\doi{10.2134/agronj1975.00021962006700040014x}.
\bibitem[{Money(1997)}]{MONEY1997173}
\bibinfo{author}{Money, N.P.}, \bibinfo{year}{1997}.
\newblock \bibinfo{title}{Wishful thinking of turgor revisited: The mechanics of fungal growth}.
\newblock \bibinfo{journal}{Fungal Genetics and Biology} \bibinfo{volume}{21}, \bibinfo{pages}{173--187}.
\newblock \URLprefix \url{https://www.sciencedirect.com/science/article/pii/S1087184597909762}, \DOIprefix\doi{10.1006/fgbi.1997.0976}.
\bibitem[{Moulton et~al.(2020)Moulton, Oliveri and Goriely}]{moulton2020multiscale}
\bibinfo{author}{Moulton, D.E.}, \bibinfo{author}{Oliveri, H.}, \bibinfo{author}{Goriely, A.}, \bibinfo{year}{2020}.
\newblock \bibinfo{title}{Multiscale integration of environmental stimuli in plant tropism produces complex behaviors}.
\newblock \bibinfo{journal}{Proceedings of the National Academy of Sciences} \bibinfo{volume}{117}, \bibinfo{pages}{32226--32237}.
\newblock \URLprefix \url{https://www.pnas.org/doi/full/10.1073/pnas.2016025117}, \DOIprefix\doi{10.1073/pnas.2016025117}.
\bibitem[{Muday(2001)}]{muday2001auxins}
\bibinfo{author}{Muday, G.K.}, \bibinfo{year}{2001}.
\newblock \bibinfo{title}{Auxins and tropisms}.
\newblock \bibinfo{journal}{Journal of plant growth regulation} \bibinfo{volume}{20}, \bibinfo{pages}{226--243}.
\newblock \URLprefix \url{https://www.proquest.com/openview/bf5663c852ddd31032ea63c0f3cadfe5/1}, \DOIprefix\doi{10.1007/s003440010027}.
\bibitem[{Murray(2003)}]{murray2mathbiol}
\bibinfo{author}{Murray, J.D.}, \bibinfo{year}{2003}.
\newblock \bibinfo{title}{Mathematical Biology II: Spatial Models and Biomedical Applications}.
\newblock Interdisciplinary Applied Mathematics. \bibinfo{edition}{3} ed., \bibinfo{publisher}{Springer}.
\newblock \URLprefix \url{https://link.springer.com/book/10.1007/b98869}, \DOIprefix\doi{10.1007/b98869}.
\bibitem[{Nakayama et~al.(2012)Nakayama, Smith, Mandel, Robinson, Kimura, Boudaoud and Kuhlemeier}]{nakayama2012mechanical}
\bibinfo{author}{Nakayama, N.}, \bibinfo{author}{Smith, R.S.}, \bibinfo{author}{Mandel, T.}, \bibinfo{author}{Robinson, S.}, \bibinfo{author}{Kimura, S.}, \bibinfo{author}{Boudaoud, A.}, \bibinfo{author}{Kuhlemeier, C.}, \bibinfo{year}{2012}.
\newblock \bibinfo{title}{Mechanical regulation of auxin-mediated growth}.
\newblock \bibinfo{journal}{Current Biology} \bibinfo{volume}{22}, \bibinfo{pages}{1468--1476}.
\newblock \URLprefix \url{https://www.cell.com/current-biology/fulltext/S0960-9822(12)00726-9}, \DOIprefix\doi{10.1016/j.cub.2012.06.050}.
\bibitem[{Needham(1934)}]{needham1934chemical}
\bibinfo{author}{Needham, J.}, \bibinfo{year}{1934}.
\newblock \bibinfo{title}{Chemical heterogony and the ground plan of animal growth}.
\newblock \bibinfo{journal}{Biological Reviews} \bibinfo{volume}{9}, \bibinfo{pages}{79--109}.
\newblock \URLprefix \url{https://onlinelibrary.wiley.com/doi/abs/10.1111/j.1469-185X.1934.tb00874.x}, \DOIprefix\doi{10.1111/j.1469-185X.1934.tb00874.x}.
\bibitem[{Newell et~al.(2008)Newell, Shipman and Sun}]{NEWELL2008421}
\bibinfo{author}{Newell, A.C.}, \bibinfo{author}{Shipman, P.D.}, \bibinfo{author}{Sun, Z.}, \bibinfo{year}{2008}.
\newblock \bibinfo{title}{Phyllotaxis: Cooperation and competition between mechanical and biochemical processes}.
\newblock \bibinfo{journal}{Journal of Theoretical Biology} \bibinfo{volume}{251}, \bibinfo{pages}{421--439}.
\newblock \URLprefix \url{https://www.sciencedirect.com/science/article/pii/S0022519307006030}, \DOIprefix\doi{10.1016/j.jtbi.2007.11.036}.
\bibitem[{Niklas and Spatz(2012)}]{niklas2012plant}
\bibinfo{author}{Niklas, K.J.}, \bibinfo{author}{Spatz, H.C.}, \bibinfo{year}{2012}.
\newblock \bibinfo{title}{Plant physics}.
\newblock \bibinfo{publisher}{University of Chicago Press}.
\newblock \URLprefix \url{https://press.uchicago.edu/ucp/books/book/chicago/P/bo12400940.html}.
\bibitem[{Nobel(2020)}]{nobel2020}
\bibinfo{author}{Nobel, P.S.}, \bibinfo{year}{2020}.
\newblock \bibinfo{title}{{Physicochemical and Environmental Plant Physiology}}.
\newblock \bibinfo{edition}{5} ed., \bibinfo{publisher}{Academic Press}.
\newblock \URLprefix \url{https://www.sciencedirect.com/book/monograph/9780128191460}, \DOIprefix\doi{10.1016/C2018-0-04662-9}.
\bibitem[{Noble(2024)}]{noble2024s}
\bibinfo{author}{Noble, D.}, \bibinfo{year}{2024}.
\newblock \bibinfo{title}{It's time to admit that genes are not the blueprint for life}.
\newblock \bibinfo{journal}{Nature} \bibinfo{volume}{626}, \bibinfo{pages}{254--255}.
\newblock \URLprefix \url{https://www.nature.com/articles/d41586-024-00327-x}, \DOIprefix\doi{10.1038/d41586-024-00327-x}.
\bibitem[{Oliveri and Cheddadi(2025)}]{oliveri2025field}
\bibinfo{author}{Oliveri, H.}, \bibinfo{author}{Cheddadi, I.}, \bibinfo{year}{2025}.
\newblock \bibinfo{title}{Hydromechanical field theory of plant morphogenesis}.
\newblock \bibinfo{journal}{Journal of the Mechanics and Physics of Solids} \bibinfo{volume}{196}, \bibinfo{pages}{106035}.
\newblock \URLprefix \url{https://www.sciencedirect.com/science/article/pii/S0022509625000110}, \DOIprefix\doi{10.1016/j.jmps.2025.106035}.
\bibitem[{Oliveri et~al.(2024)Oliveri, Moulton, Harrington and Goriely}]{Oliveri2024active}
\bibinfo{author}{Oliveri, H.}, \bibinfo{author}{Moulton, D.E.}, \bibinfo{author}{Harrington, H.A.}, \bibinfo{author}{Goriely, A.}, \bibinfo{year}{2024}.
\newblock \bibinfo{title}{Active shape control by plants in dynamic environments}.
\newblock \bibinfo{journal}{Physical Review E} \bibinfo{volume}{110}, \bibinfo{pages}{014405}.
\newblock \URLprefix \url{https://link.aps.org/doi/10.1103/PhysRevE.110.014405}, \DOIprefix\doi{10.1103/PhysRevE.110.014405}.
\bibitem[{Oliveri et~al.(2018)Oliveri, Traas, Godin and Ali}]{oliveri2019regulation}
\bibinfo{author}{Oliveri, H.}, \bibinfo{author}{Traas, J.}, \bibinfo{author}{Godin, C.}, \bibinfo{author}{Ali, O.}, \bibinfo{year}{2018}.
\newblock \bibinfo{title}{Regulation of plant cell wall stiffness by mechanical stress: a mesoscale physical model}.
\newblock \bibinfo{journal}{Journal of mathematical biology} \bibinfo{volume}{78}, \bibinfo{pages}{625--653}.
\newblock \URLprefix \url{https://link.springer.com/article/10.1007/s00285-018-1286-y}, \DOIprefix\doi{10.1007/s00285-018-1286-y}.
\bibitem[{Ortega(1985)}]{ortega_augmented_1985}
\bibinfo{author}{Ortega, J.K.E.}, \bibinfo{year}{1985}.
\newblock \bibinfo{title}{Augmented {Growth} {Equation} for {Cell} {Wall} {Expansion}}.
\newblock \bibinfo{journal}{Plant Physiology} \bibinfo{volume}{79}, \bibinfo{pages}{318--320}.
\newblock \URLprefix \url{http://www.plantphysiol.org/content/79/1/318}, \DOIprefix\doi{10.1104/pp.79.1.318}.
\bibitem[{Paredez et~al.(2006)Paredez, Somerville and Ehrhardt}]{paredez2006visualization}
\bibinfo{author}{Paredez, A.R.}, \bibinfo{author}{Somerville, C.R.}, \bibinfo{author}{Ehrhardt, D.W.}, \bibinfo{year}{2006}.
\newblock \bibinfo{title}{Visualization of cellulose synthase demonstrates functional association with microtubules}.
\newblock \bibinfo{journal}{Science} \bibinfo{volume}{312}, \bibinfo{pages}{1491--1495}.
\newblock \URLprefix \url{https://www.science.org/doi/10.1126/science.1126551}, \DOIprefix\doi{10.1126/science.1126551}.
\bibitem[{Passioura and Boyer(2003)}]{passioura2003tissue}
\bibinfo{author}{Passioura, J.B.}, \bibinfo{author}{Boyer, J.S.}, \bibinfo{year}{2003}.
\newblock \bibinfo{title}{Tissue stresses and resistance to water flow conspire to uncouple the water potential of the epidermis from that of the xylem in elongating plant stems}.
\newblock \bibinfo{journal}{Functional Plant Biology} \bibinfo{volume}{30}, \bibinfo{pages}{325--334}.
\newblock \URLprefix \url{https://www.publish.csiro.au/fp/FP02202}, \DOIprefix\doi{10.1071/FP02202}.
\bibitem[{Peters and Tomos(1996)}]{peters1996history}
\bibinfo{author}{Peters, W.}, \bibinfo{author}{Tomos, A.}, \bibinfo{year}{1996}.
\newblock \bibinfo{title}{The history of tissue tension}.
\newblock \bibinfo{journal}{Annals of Botany} \bibinfo{volume}{77}, \bibinfo{pages}{657--665}.
\newblock \URLprefix \url{https://academic.oup.com/aob/article/77/6/657/2389842}, \DOIprefix\doi{10.1006/anbo.1996.0082}.
\bibitem[{P{\'e}zard(1918)}]{pezard1918conditionnement}
\bibinfo{author}{P{\'e}zard, A.}, \bibinfo{year}{1918}.
\newblock \bibinfo{title}{Le conditionnement physiologique des caract{\`e}res sexuels secondaires chez les oiseaux}.
\newblock \bibinfo{journal}{Bulletin Biologique de la France et de la Belgique} \bibinfo{volume}{52}, \bibinfo{pages}{1--176}.
\bibitem[{Philip(1958)}]{philip1958propagation}
\bibinfo{author}{Philip, J.}, \bibinfo{year}{1958}.
\newblock \bibinfo{title}{Propagation of turgor and other properties through cell aggregations.}
\newblock \bibinfo{journal}{Plant Physiology} \bibinfo{volume}{33}, \bibinfo{pages}{271}.
\newblock \URLprefix \url{https://academic.oup.com/plphys/article/33/4/271/6089488}, \DOIprefix\doi{10.1104/pp.33.4.271}.
\bibitem[{Pietruszka and Lewicka(2007)}]{pietruszka2007anisotropic}
\bibinfo{author}{Pietruszka, M.}, \bibinfo{author}{Lewicka, S.}, \bibinfo{year}{2007}.
\newblock \bibinfo{title}{Anisotropic plant growth due to phototropism}.
\newblock \bibinfo{journal}{Journal of mathematical biology} \bibinfo{volume}{54}, \bibinfo{pages}{45--55}.
\newblock \URLprefix \url{https://link.springer.com/article/10.1007/s00285-006-0045-7}, \DOIprefix\doi{10.1007/s00285-006-0045-7}.
\bibitem[{Plant(1982)}]{plant1982continuum}
\bibinfo{author}{Plant, R.E.}, \bibinfo{year}{1982}.
\newblock \bibinfo{title}{A continuum model for root growth}.
\newblock \bibinfo{journal}{Journal of Theoretical Biology} \bibinfo{volume}{98}, \bibinfo{pages}{45--59}.
\newblock \URLprefix \url{https://www.sciencedirect.com/science/article/pii/0022519382900571}, \DOIprefix\doi{10.1016/0022-5193(82)90057-1}.
\bibitem[{Ramos et~al.(2021)Ramos, Maizel and Alim}]{ramos2021tissue}
\bibinfo{author}{Ramos, J.R.D.}, \bibinfo{author}{Maizel, A.}, \bibinfo{author}{Alim, K.}, \bibinfo{year}{2021}.
\newblock \bibinfo{title}{Tissue-wide integration of mechanical cues promotes effective auxin patterning}.
\newblock \bibinfo{journal}{The European Physical Journal Plus} \bibinfo{volume}{136}, \bibinfo{pages}{250}.
\newblock \URLprefix \url{https://link.springer.com/article/10.1140/epjp/s13360-021-01204-6}, \DOIprefix\doi{10.1140/epjp/s13360-021-01204-6}.
\bibitem[{Ray et~al.(1972)Ray, Green and Cleland}]{ray_role_1972}
\bibinfo{author}{Ray, P.M.}, \bibinfo{author}{Green, P.B.}, \bibinfo{author}{Cleland, R.}, \bibinfo{year}{1972}.
\newblock \bibinfo{title}{Role of {Turgor} in {Plant} {Cell} {Growth}}.
\newblock \bibinfo{journal}{Nature} \bibinfo{volume}{239}, \bibinfo{pages}{163--164}.
\newblock \URLprefix \url{http://www.nature.com/nature/journal/v239/n5368/pdf/239163a0.pdf}, \DOIprefix\doi{10.1038/239163a0}.
\bibitem[{Richards and Kavanagh(1943)}]{richards1943analysis}
\bibinfo{author}{Richards, O.W.}, \bibinfo{author}{Kavanagh, A.J.}, \bibinfo{year}{1943}.
\newblock \bibinfo{title}{The analysis of the relative growth gradients and changing form of growing organisms: illustrated by the tobacco leaf}.
\newblock \bibinfo{journal}{The American Naturalist} \bibinfo{volume}{77}, \bibinfo{pages}{385--399}.
\newblock \URLprefix \url{https://www.journals.uchicago.edu/doi/10.1086/281140}, \DOIprefix\doi{10.1086/281140}.
\bibitem[{Rodriguez et~al.(1994)Rodriguez, Hoger and McCulloch}]{rodriguez1994stress}
\bibinfo{author}{Rodriguez, E.K.}, \bibinfo{author}{Hoger, A.}, \bibinfo{author}{McCulloch, A.D.}, \bibinfo{year}{1994}.
\newblock \bibinfo{title}{{Stress-dependent finite growth in soft elastic tissues}}.
\newblock \bibinfo{journal}{Journal of Biomechanics} \bibinfo{volume}{27}, \bibinfo{pages}{455--467}.
\newblock \URLprefix \url{https://www.sciencedirect.com/science/article/pii/0021929094900213}, \DOIprefix\doi{10.1016/0021-9290(94)90021-3}.
\bibitem[{Rolland-Lagan et~al.(2005)Rolland-Lagan, Coen, Impey and Bangham}]{Rolland-Lagan.2005}
\bibinfo{author}{Rolland-Lagan, A.G.}, \bibinfo{author}{Coen, E.}, \bibinfo{author}{Impey, S.J.}, \bibinfo{author}{Bangham, J.A.}, \bibinfo{year}{2005}.
\newblock \bibinfo{title}{{A computational method for inferring growth parameters and shape changes during development based on clonal analysis}}.
\newblock \bibinfo{journal}{Journal of Theoretical Biology} \bibinfo{volume}{232}, \bibinfo{pages}{157 -- 177}.
\newblock \URLprefix \url{https://www.sciencedirect.com/science/article/pii/S0022519304003054}, \DOIprefix\doi{10.1016/j.jtbi.2004.04.045}.
\bibitem[{Ruan et~al.(2001)Ruan, Llewellyn and Furbank}]{ruan2001}
\bibinfo{author}{Ruan, Y.L.}, \bibinfo{author}{Llewellyn, D.J.}, \bibinfo{author}{Furbank, R.T.}, \bibinfo{year}{2001}.
\newblock \bibinfo{title}{The control of single-celled cotton fiber elongation by developmentally reversible gating of plasmodesmata and coordinated expression of sucrose and {K+} transporters and expansin}.
\newblock \bibinfo{journal}{The Plant Cell} \bibinfo{volume}{13}, \bibinfo{pages}{47--60}.
\newblock \URLprefix \url{https://academic.oup.com/plcell/article/13/1/47/6009433}, \DOIprefix\doi{10.1105/tpc.13.1.47}.
\bibitem[{Rudge and Haseloff(2005)}]{rudge2005computational}
\bibinfo{author}{Rudge, T.}, \bibinfo{author}{Haseloff, J.}, \bibinfo{year}{2005}.
\newblock \bibinfo{title}{A computational model of cellular morphogenesis in plants}, in: \bibinfo{booktitle}{European Conference on Artificial Life}, \bibinfo{organization}{Springer}. pp. \bibinfo{pages}{78--87}.
\newblock \URLprefix \url{https://link.springer.com/chapter/10.1007/11553090_9}, \DOIprefix\doi{10.1007/11553090_9}.
\bibitem[{Rueda-Contreras and Arag{\'o}n(2014)}]{rueda2014alan}
\bibinfo{author}{Rueda-Contreras, M.}, \bibinfo{author}{Arag{\'o}n, J.}, \bibinfo{year}{2014}.
\newblock \bibinfo{title}{{Alan Turing's chemical theory of phyllotaxis}}.
\newblock \bibinfo{journal}{Revista mexicana de f{\'\i}sica E} \bibinfo{volume}{60}, \bibinfo{pages}{1--12}.
\newblock \URLprefix \url{https://www.scielo.org.mx/scielo.php?pid=S1870-35422014000100001&script=sci_arttext}.
\bibitem[{Rueda-Contreras et~al.(2018)Rueda-Contreras, Romero-Arias, Aragon and Barrio}]{rueda2018curvature}
\bibinfo{author}{Rueda-Contreras, M.D.}, \bibinfo{author}{Romero-Arias, J.R.}, \bibinfo{author}{Aragon, J.L.}, \bibinfo{author}{Barrio, R.A.}, \bibinfo{year}{2018}.
\newblock \bibinfo{title}{Curvature-driven spatial patterns in growing 3d domains: A mechanochemical model for phyllotaxis}.
\newblock \bibinfo{journal}{PloS one} \bibinfo{volume}{13}, \bibinfo{pages}{e0201746}.
\newblock \URLprefix \url{https://journals.plos.org/plosone/article?id=10.1371/journal.pone.0201746}, \DOIprefix\doi{10.1371/journal.pone.0201746}.
\bibitem[{Sachs(1882)}]{sachs1882text}
\bibinfo{author}{Sachs, J.}, \bibinfo{year}{1882}.
\newblock \bibinfo{title}{Text-book of botany: morphological and physiological}.
\newblock \bibinfo{edition}{2} ed., \bibinfo{publisher}{Clarendon Press}, \bibinfo{address}{Oxford}.
\newblock \URLprefix \url{https://www.biodiversitylibrary.org/bibliography/58583}, \DOIprefix\doi{10.5962/bhl.title.58583}.
\bibitem[{Sassi et~al.(2014)Sassi, Ali, Boudon, Cloarec, Abad, Cellier, Chen, Gilles, Milani, Friml, Vernoux, Godin, Hamant and Traas}]{sassi_auxin-mediated_2014}
\bibinfo{author}{Sassi, M.}, \bibinfo{author}{Ali, O.}, \bibinfo{author}{Boudon, F.}, \bibinfo{author}{Cloarec, G.}, \bibinfo{author}{Abad, U.}, \bibinfo{author}{Cellier, C.}, \bibinfo{author}{Chen, X.}, \bibinfo{author}{Gilles, B.}, \bibinfo{author}{Milani, P.}, \bibinfo{author}{Friml, J.}, \bibinfo{author}{Vernoux, T.}, \bibinfo{author}{Godin, C.}, \bibinfo{author}{Hamant, O.}, \bibinfo{author}{Traas, J.}, \bibinfo{year}{2014}.
\newblock \bibinfo{title}{An {Auxin}-{Mediated} {Shift} toward {Growth} {Isotropy} {Promotes} {Organ} {Formation} at the {Shoot} {Meristem} in {Arabidopsis}}.
\newblock \bibinfo{journal}{Current Biology} \bibinfo{volume}{24}, \bibinfo{pages}{2335--2342}.
\newblock \URLprefix \url{http://linkinghub.elsevier.com/retrieve/pii/S0960982214010495}, \DOIprefix\doi{10.1016/j.cub.2014.08.036}.
\bibitem[{Saw et~al.(2018)Saw, Xi, Ladoux and Lim}]{saw2018biological}
\bibinfo{author}{Saw, T.B.}, \bibinfo{author}{Xi, W.}, \bibinfo{author}{Ladoux, B.}, \bibinfo{author}{Lim, C.T.}, \bibinfo{year}{2018}.
\newblock \bibinfo{title}{Biological tissues as active nematic liquid crystals}.
\newblock \bibinfo{journal}{Advanced materials} \bibinfo{volume}{30}, \bibinfo{pages}{1802579}.
\newblock \URLprefix \url{https://advanced.onlinelibrary.wiley.com/doi/full/10.1002/adma.201802579}, \DOIprefix\doi{10.1002/adma.201802579}.
\bibitem[{Scarpella et~al.(2010)Scarpella, Barkoulas and Tsiantis}]{scarpella2010control}
\bibinfo{author}{Scarpella, E.}, \bibinfo{author}{Barkoulas, M.}, \bibinfo{author}{Tsiantis, M.}, \bibinfo{year}{2010}.
\newblock \bibinfo{title}{Control of leaf and vein development by auxin}.
\newblock \bibinfo{journal}{Cold Spring Harbor perspectives in biology} \bibinfo{volume}{2}, \bibinfo{pages}{a001511}.
\newblock \URLprefix \url{https://cshperspectives.cshlp.org/content/2/1/a001511.long}, \DOIprefix\doi{10.1101/cshperspect.a001511}.
\bibitem[{Schwendener(1878)}]{schwendener1878mechanische}
\bibinfo{author}{Schwendener, S.}, \bibinfo{year}{1878}.
\newblock \bibinfo{title}{{Mechanische Theorie der Blattstellungen}}.
\newblock \bibinfo{publisher}{W. Engelmann}, \bibinfo{address}{Leipzig}.
\bibitem[{Sharon and Efrati(2010)}]{sharon2010mechanics}
\bibinfo{author}{Sharon, E.}, \bibinfo{author}{Efrati, E.}, \bibinfo{year}{2010}.
\newblock \bibinfo{title}{The mechanics of non-euclidean plates}.
\newblock \bibinfo{journal}{Soft Matter} \bibinfo{volume}{6}, \bibinfo{pages}{5693--5704}.
\newblock \URLprefix \url{https://pubs.rsc.org/en/content/articlehtml/2010/sm/c0sm00479k}, \DOIprefix\doi{10.1039/C0SM00479K}.
\bibitem[{Silk(1984)}]{silk1984quantitative}
\bibinfo{author}{Silk, W.K.}, \bibinfo{year}{1984}.
\newblock \bibinfo{title}{Quantitative descriptions of development}.
\newblock \bibinfo{journal}{Annual Review of Plant Physiology} \bibinfo{volume}{35}, \bibinfo{pages}{479--518}.
\newblock \URLprefix \url{https://www.annualreviews.org/content/journals/10.1146/annurev.pp.35.060184.002403}, \DOIprefix\doi{10.1146/annurev.pp.35.060184.002403}.
\bibitem[{Silk and Erickson(1979)}]{SILK1979481}
\bibinfo{author}{Silk, W.K.}, \bibinfo{author}{Erickson, R.O.}, \bibinfo{year}{1979}.
\newblock \bibinfo{title}{Kinematics of plant growth}.
\newblock \bibinfo{journal}{Journal of Theoretical Biology} \bibinfo{volume}{76}, \bibinfo{pages}{481--501}.
\newblock \URLprefix \url{https://www.sciencedirect.com/science/article/pii/0022519379900146}, \DOIprefix\doi{10.1016/0022-5193(79)90014-6}.
\bibitem[{Silk and Wagner(1980)}]{silk1980growth}
\bibinfo{author}{Silk, W.K.}, \bibinfo{author}{Wagner, K.K.}, \bibinfo{year}{1980}.
\newblock \bibinfo{title}{Growth-sustaining water potential distributions in the primary corn root: A noncompartmented continuum model}.
\newblock \bibinfo{journal}{Plant Physiology} \bibinfo{volume}{66}, \bibinfo{pages}{859--863}.
\newblock \URLprefix \url{https://academic.oup.com/plphys/article/66/5/859/6077210}, \DOIprefix\doi{10.1104/pp.66.5.859}.
\bibitem[{Silveira et~al.(2025)Silveira, Collet, Haque, Lapierre, Bagniewska-Zadworna, Smith, Gosselin, Routier-Kierzkowska and Kierzkowski}]{silveira2025mechanical}
\bibinfo{author}{Silveira, S.R.}, \bibinfo{author}{Collet, L.}, \bibinfo{author}{Haque, S.M.}, \bibinfo{author}{Lapierre, L.}, \bibinfo{author}{Bagniewska-Zadworna, A.}, \bibinfo{author}{Smith, R.S.}, \bibinfo{author}{Gosselin, F.P.}, \bibinfo{author}{Routier-Kierzkowska, A.L.}, \bibinfo{author}{Kierzkowski, D.}, \bibinfo{year}{2025}.
\newblock \bibinfo{title}{Mechanical interactions between tissue layers underlie plant morphogenesis}.
\newblock \bibinfo{journal}{Nature Plants} \bibinfo{volume}{11}, \bibinfo{pages}{909--923}.
\newblock \URLprefix \url{https://www.nature.com/articles/s41477-025-01944-8}, \DOIprefix\doi{10.1038/s41477-025-01944-8}.
\bibitem[{Sinnott(1939)}]{sinnott1939cell}
\bibinfo{author}{Sinnott, E.W.}, \bibinfo{year}{1939}.
\newblock \bibinfo{title}{The cell-organ relationship in plant organization}.
\newblock \bibinfo{journal}{Growth} \bibinfo{volume}{3}, \bibinfo{pages}{77--86}.
\bibitem[{Skalak et~al.(1982)Skalak, Dasgupta, Moss, Otten, Dullemeijer and Vilmann}]{skalak1982analytical}
\bibinfo{author}{Skalak, R.}, \bibinfo{author}{Dasgupta, G.}, \bibinfo{author}{Moss, M.}, \bibinfo{author}{Otten, E.}, \bibinfo{author}{Dullemeijer, P.}, \bibinfo{author}{Vilmann, H.}, \bibinfo{year}{1982}.
\newblock \bibinfo{title}{Analytical description of growth}.
\newblock \bibinfo{journal}{Journal of theoretical biology} \bibinfo{volume}{94}, \bibinfo{pages}{555--577}.
\newblock \URLprefix \url{https://www.sciencedirect.com/science/article/abs/pii/0022519382903010}, \DOIprefix\doi{https://doi.org/10.1016/0022-5193(82)90301-0}.
\bibitem[{Smith et~al.(2006a)Smith, Guyomarc'h, Mandel, Reinhardt, Kuhlemeier and Prusinkiewicz}]{smith_plausible_2006}
\bibinfo{author}{Smith, R.S.}, \bibinfo{author}{Guyomarc'h, S.}, \bibinfo{author}{Mandel, T.}, \bibinfo{author}{Reinhardt, D.}, \bibinfo{author}{Kuhlemeier, C.}, \bibinfo{author}{Prusinkiewicz, P.}, \bibinfo{year}{2006}a.
\newblock \bibinfo{title}{A plausible model of phyllotaxis}.
\newblock \bibinfo{journal}{Proceedings of the National Academy of Sciences} \bibinfo{volume}{103}, \bibinfo{pages}{1301--1306}.
\newblock \URLprefix \url{http://www.pnas.org/content/103/5/1301}, \DOIprefix\doi{10.1073/pnas.0510457103}.
\bibitem[{Smith et~al.(2006b)Smith, Kuhlemeier and Prusinkiewicz}]{smith2006inhibition}
\bibinfo{author}{Smith, R.S.}, \bibinfo{author}{Kuhlemeier, C.}, \bibinfo{author}{Prusinkiewicz, P.}, \bibinfo{year}{2006}b.
\newblock \bibinfo{title}{Inhibition fields for phyllotactic pattern formation: a simulation study}.
\newblock \bibinfo{journal}{Canadian Journal of Botany} \bibinfo{volume}{84}, \bibinfo{pages}{1635 -- 1649}.
\newblock \URLprefix \url{http://www.nrcresearchpress.com/doi/abs/10.1139/b06-133}, \DOIprefix\doi{10.1139/b06-133}.
\bibitem[{Smithers et~al.(2024)Smithers, Luo and Dyson}]{smithers2024continuum}
\bibinfo{author}{Smithers, E.T.}, \bibinfo{author}{Luo, J.}, \bibinfo{author}{Dyson, R.J.}, \bibinfo{year}{2024}.
\newblock \bibinfo{title}{A continuum mechanics model of the plant cell wall reveals interplay between enzyme action and cell wall structure}.
\newblock \bibinfo{journal}{The European Physical Journal E} \bibinfo{volume}{47}, \bibinfo{pages}{1}.
\newblock \URLprefix \url{https://link.springer.com/article/10.1140/epje/s10189-023-00396-2}, \DOIprefix\doi{10.1140/epje/s10189-023-00396-2}.
\bibitem[{Swinton(2004)}]{swinton2004watching}
\bibinfo{author}{Swinton, J.}, \bibinfo{year}{2004}.
\newblock \bibinfo{title}{{Watching the daisies grow: Turing and Fibonacci phyllotaxis}}, in: \bibinfo{editor}{Teuscher, C.} (Ed.), \bibinfo{booktitle}{Alan Turing: life and legacy of a great thinker}. \bibinfo{publisher}{Springer}, pp. \bibinfo{pages}{477--498}.
\newblock \URLprefix \url{https://link.springer.com/chapter/10.1007/978-3-662-05642-4_20}, \DOIprefix\doi{10.1007/978-3-662-05642-4_20}.
\bibitem[{Thompson(1917)}]{thompson1917arcy}
\bibinfo{author}{Thompson, D.W.}, \bibinfo{year}{1917}.
\newblock \bibinfo{title}{On growth and form}.
\newblock \bibinfo{edition}{1} ed., \bibinfo{publisher}{Cambridge University Press}, \bibinfo{address}{Cambridge}.
\bibitem[{Toner(2024)}]{Toner_2024}
\bibinfo{author}{Toner, J.}, \bibinfo{year}{2024}.
\newblock \bibinfo{title}{The Physics of Flocking: Birth, Death, and Flight in Active Matter}.
\newblock \bibinfo{publisher}{Cambridge University Press}, \bibinfo{address}{Cambridge}.
\newblock \URLprefix \url{https://www.cambridge.org/core/books/physics-of-flocking/7156611CA514710F371ECCCB07558976}, \DOIprefix\doi{10.1017/9781108993623}.
\bibitem[{Traas(2013)}]{traas_phyllotaxis_2013}
\bibinfo{author}{Traas, J.}, \bibinfo{year}{2013}.
\newblock \bibinfo{title}{Phyllotaxis}.
\newblock \bibinfo{journal}{Development} \bibinfo{volume}{140}, \bibinfo{pages}{249--253}.
\newblock \URLprefix \url{http://dev.biologists.org/content/140/2/249}, \DOIprefix\doi{10.1242/dev.074740}.
\bibitem[{Tsugawa et~al.(2017)Tsugawa, Hervieux, Kierzkowski, Routier-Kierzkowska, Sapala, Hamant, Smith, Roeder, Boudaoud and Li}]{tsugawa2017clones}
\bibinfo{author}{Tsugawa, S.}, \bibinfo{author}{Hervieux, N.}, \bibinfo{author}{Kierzkowski, D.}, \bibinfo{author}{Routier-Kierzkowska, A.L.}, \bibinfo{author}{Sapala, A.}, \bibinfo{author}{Hamant, O.}, \bibinfo{author}{Smith, R.S.}, \bibinfo{author}{Roeder, A.H.}, \bibinfo{author}{Boudaoud, A.}, \bibinfo{author}{Li, C.B.}, \bibinfo{year}{2017}.
\newblock \bibinfo{title}{Clones of cells switch from reduction to enhancement of size variability in \textit{Arabidopsis} sepals}.
\newblock \bibinfo{journal}{Development} \bibinfo{volume}{144}, \bibinfo{pages}{4398--4405}.
\newblock \URLprefix \url{https://journals.biologists.com/dev/article/144/23/4398/19276}, \DOIprefix\doi{10.1242/dev.153999}.
\bibitem[{Turing(1952)}]{turing1952chemical}
\bibinfo{author}{Turing, A.M.}, \bibinfo{year}{1952}.
\newblock \bibinfo{title}{The chemical basis of morphogenesis}.
\newblock \bibinfo{journal}{Philosophical Transactions of the Royal Society of London B} \bibinfo{volume}{237}, \bibinfo{pages}{37--72}.
\newblock \URLprefix \url{https://royalsocietypublishing.org/doi/10.1098/rstb.1952.0012}, \DOIprefix\doi{10.1098/rstb.1952.0012}.
\bibitem[{Vandiver and Goriely(2009)}]{vandiver2009morpho}
\bibinfo{author}{Vandiver, R.}, \bibinfo{author}{Goriely, A.}, \bibinfo{year}{2009}.
\newblock \bibinfo{title}{Morpho-elastodynamics: the long-time dynamics of elastic growth}.
\newblock \bibinfo{journal}{Journal of biological dynamics} \bibinfo{volume}{3}, \bibinfo{pages}{180--195}.
\newblock \URLprefix \url{https://www.tandfonline.com/doi/full/10.1080/17513750802304885}, \DOIprefix\doi{10.1080/17513750802304885}.
\bibitem[{Vernoux et~al.(2021)Vernoux, Besnard and Godin}]{Vernoux.2021}
\bibinfo{author}{Vernoux, T.}, \bibinfo{author}{Besnard, F.}, \bibinfo{author}{Godin, C.}, \bibinfo{year}{2021}.
\newblock \bibinfo{title}{{What shoots can teach about theories of plant form}}.
\newblock \bibinfo{journal}{Nature Plants} \bibinfo{volume}{7}, \bibinfo{pages}{716–724}.
\newblock \URLprefix \url{https://www.nature.com/articles/s41477-021-00930-0}, \DOIprefix\doi{10.1038/s41477-021-00930-0}.
\bibitem[{Vernoux et~al.(2010)Vernoux, Besnard and Traas}]{vernoux_auxin_2010}
\bibinfo{author}{Vernoux, T.}, \bibinfo{author}{Besnard, F.}, \bibinfo{author}{Traas, J.}, \bibinfo{year}{2010}.
\newblock \bibinfo{title}{Auxin at the {Shoot} {Apical} {Meristem}}.
\newblock \bibinfo{journal}{Cold Spring Harbor Perspectives in Biology} \bibinfo{volume}{2}, \bibinfo{pages}{a001487}.
\newblock \URLprefix \url{http://cshperspectives.cshlp.org/content/2/4/a001487}, \DOIprefix\doi{10.1101/cshperspect.a001487}.
\bibitem[{Whippo and Hangarter(2006)}]{whippo2006phototropism}
\bibinfo{author}{Whippo, C.W.}, \bibinfo{author}{Hangarter, R.P.}, \bibinfo{year}{2006}.
\newblock \bibinfo{title}{Phototropism: bending towards enlightenment}.
\newblock \bibinfo{journal}{The Plant Cell} \bibinfo{volume}{18}, \bibinfo{pages}{1110--1119}.
\newblock \URLprefix \url{https://academic.oup.com/plcell/article/18/5/1110/6114902}, \DOIprefix\doi{10.1105/tpc.105.039669}.
\bibitem[{Wiegers et~al.(2009)Wiegers, Cheer and Silk}]{wiegers2009modeling}
\bibinfo{author}{Wiegers, B.S.}, \bibinfo{author}{Cheer, A.Y.}, \bibinfo{author}{Silk, W.K.}, \bibinfo{year}{2009}.
\newblock \bibinfo{title}{Modeling the hydraulics of root growth in three dimensions with phloem water sources}.
\newblock \bibinfo{journal}{Plant Physiology} \bibinfo{volume}{150}, \bibinfo{pages}{2092--2103}.
\newblock \URLprefix \url{https://academic.oup.com/plphys/article/150/4/2092/6108078?login=true}, \DOIprefix\doi{10.1104/pp.109.138198}.
\bibitem[{Wolf et~al.(1986)Wolf, Silk and Plant}]{wolf1986quantitative}
\bibinfo{author}{Wolf, S.D.}, \bibinfo{author}{Silk, W.K.}, \bibinfo{author}{Plant, R.E.}, \bibinfo{year}{1986}.
\newblock \bibinfo{title}{Quantitative patterns of leaf expansion: comparison of normal and malformed leaf growth in \textit{Vitis vinifera} cv. ruby red}.
\newblock \bibinfo{journal}{American Journal of Botany} \bibinfo{volume}{73}, \bibinfo{pages}{832--846}.
\newblock \URLprefix \url{https://bsapubs.onlinelibrary.wiley.com/doi/abs/10.1002/j.1537-2197.1986.tb12121.x}, \DOIprefix\doi{10.1002/j.1537-2197.1986.tb12121.x}.
\bibitem[{Yavari(2013)}]{yavari2013compatibility}
\bibinfo{author}{Yavari, A.}, \bibinfo{year}{2013}.
\newblock \bibinfo{title}{Compatibility equations of nonlinear elasticity for non-simply-connected bodies}.
\newblock \bibinfo{journal}{Archive for Rational Mechanics and Analysis} \bibinfo{volume}{209}, \bibinfo{pages}{237--253}.
\newblock \URLprefix \url{https://link.springer.com/article/10.1007/s00205-013-0621-0}, \DOIprefix\doi{10.1007/s00205-013-0621-0}.
\bibitem[{Zhang et~al.(2024)Zhang, Ramakanth and Long}]{zhang2024biomechanics}
\bibinfo{author}{Zhang, X.}, \bibinfo{author}{Ramakanth, K.K.}, \bibinfo{author}{Long, Y.}, \bibinfo{year}{2024}.
\newblock \bibinfo{title}{The biomechanics of turgor pressure}.
\newblock \bibinfo{journal}{Current Biology} \bibinfo{volume}{34}, \bibinfo{pages}{R986--R991}.
\newblock \URLprefix \url{https://www.cell.com/current-biology/fulltext/S0960-9822(24)00916-3}, \DOIprefix\doi{10.1016/j.cub.2024.07.013}.
\bibitem[{Zhang et~al.(2025)Zhang, Cohen, Moshe and Sharon}]{zhang2025geometrically}
\bibinfo{author}{Zhang, Y.}, \bibinfo{author}{Cohen, O.Y.}, \bibinfo{author}{Moshe, M.}, \bibinfo{author}{Sharon, E.}, \bibinfo{year}{2025}.
\newblock \bibinfo{title}{Geometrically frustrated rose petals}.
\newblock \bibinfo{journal}{Science} \bibinfo{volume}{388}, \bibinfo{pages}{520--524}.
\newblock \URLprefix \url{https://www.science.org/doi/abs/10.1126/science.adt0672}, \DOIprefix\doi{10.1126/science.adt0672}.
\bibitem[{Zhao et~al.(2020)Zhao, Du, Oliveri, Zhou, Ali, Chen, Feng, Wang, L{\"u}, Long, Schneider, Sampathkumar, Godin and Traas}]{zhao2020microtubule}
\bibinfo{author}{Zhao, F.}, \bibinfo{author}{Du, F.}, \bibinfo{author}{Oliveri, H.}, \bibinfo{author}{Zhou, L.}, \bibinfo{author}{Ali, O.}, \bibinfo{author}{Chen, W.}, \bibinfo{author}{Feng, S.}, \bibinfo{author}{Wang, Q.}, \bibinfo{author}{L{\"u}, S.}, \bibinfo{author}{Long, M.}, \bibinfo{author}{Schneider, R.}, \bibinfo{author}{Sampathkumar, A.}, \bibinfo{author}{Godin, C.}, \bibinfo{author}{Traas, J.}, \bibinfo{year}{2020}.
\newblock \bibinfo{title}{Microtubule-mediated wall anisotropy contributes to leaf blade flattening}.
\newblock \bibinfo{journal}{Current Biology} \bibinfo{volume}{30}, \bibinfo{pages}{3972--3985}.
\newblock \URLprefix \url{https://www.sciencedirect.com/science/article/pii/S0960982220311015}, \DOIprefix\doi{10.1016/j.cub.2020.07.076}.

\end{thebibliography}

\end{document}